\documentclass[aps,prl,superscriptaddress,twocolumn,showpacs,preprintnumbers,amsmath,amssymb]{revtex4}

\usepackage{amsopn}
\usepackage{braket}

\usepackage{amsmath}
\usepackage{mathrsfs}
\usepackage{dsfont}
\usepackage{amstext}
\usepackage{amssymb}
\usepackage{amsbsy}
\usepackage{bbm}
\usepackage{amsthm}
\usepackage{graphicx}
\usepackage{color}
\usepackage[colorlinks,citecolor=blue]{hyperref}

\newcommand{\ie}{{\emph{i.e.~}}}
\newcommand{\Ref}[1]{Ref.~\onlinecite{#1}}

\renewcommand{\ie}{{\emph{i.e.~}}} \makeatletter

\newcommand{\Rmnum}[1]{\expandafter\@slowromancap\romannumeral #1@}
\makeatother
\newcommand{\imth}{\hspace{1pt}\mathrm{i}\hspace{1pt}}

\newcommand{\tk}{{\textbf{k}}}

\begin{document}

\title{Majorana fermions in spin-singlet nodal superconductors with coexisting non-collinear magnetic order}
\author{Yuan-Ming Lu}
\affiliation{Department of Physics, University of California, Berkeley, CA 94720}
\affiliation{Materials Science Division, Lawrence Berkeley National Laboratories, Berkeley, CA 94720}

\author{Ziqiang Wang}
\affiliation{{Department of Physics, Boston College, Chestnut
Hill, Massachusetts 02467}}
\date{{\small \today}}


\begin{abstract}
Realizations of Majorana fermions in solid state materials have attracted great interests recently in connection to topological order and quantum information processing. We propose a novel way to create Majorana fermions in superconductors. We show that an incipient non-collinear magnetic order turns a spin-singlet superconductor with nodes into a topological superconductor with a stable Majorana bound state in the vortex core; at a topologically-stable magnetic point defect; and on the edge. We argue that such an exotic non-Abelian phase can be realized in extended $t$-$J$ models on the triangular and square lattices. It is promising to search for Majorana fermions in correlated electron materials where nodal superconductivity and magnetism are two common caricatures.
\end{abstract}


\maketitle

A Majorana fermion is an electrically neutral fermion whose antiparticle is itself \cite{Wilczek2009}.
In recent years, Majorana fermions have attracted growing attention in condensed matter physics \cite{Nayak2008,Hasan2010,Qi2011}. Specifically, Majorana fermions can be realized as zero-energy bound states in the vortex core or on the edge of certain two-dimensional superconductors. Instead of the usual Bose or Fermi statistics, these vortices obey non-Abelian statistics \cite{Moore1991,Nayak1996,Fradkin1998,Read2000,Ivanov2001,Stone2006} as a manifestation of topological order \cite{Wen1990b,Wen1991b}. Due to this remarkable feature,  Majorana bound states (MBSs) can be utilized for topologically-protected qubits in fault-tolerant quantum computation \cite{Kitaev2003,DasSarma2005,Nayak2008}. Several systems have been proposed to realize MBSs, such as even-denominator fractional quantum Hall states \cite{Moore1991,Nayak1996,Read2000,Lu2010c}, $p+\imth p$ superconductors \cite{Read2000,Ivanov2001,Stone2006} and superfluids \cite{Gurarie2005,Cooper2009}, superconductor-topological insulator interfaces \cite{Fu2008,Stanescu2010,Linder2010,Weng2011}, $s$-wave Rashba superconductor \cite{Sato2009,Sau2010,Alicea2010} and spin-orbit-coupled nodal superconductors\cite{Sato2010}.

In this work, we present a novel realization of MBSs in spin-singlet superconductors with nodal excitations.We show that when a coexisting non-collinear magnetic order (NCMO) develops with a wavevector connecting two nodes at opposite momenta, there will be one MBS in each vortex core and on the edge of such topological superconductors. Moreover, each stable point defect of the NCMO also hosts a MBS. We demonstrate our proposal with two explicit examples. The first one is a nodal $d+\imth d$ superconductor \cite{Zhou2008} coexisting with $1\times3$ (or $3\times3$) coplanar magnetic order on the triangular lattice. We argue that this state is likely to be realized in a doped $t$-$J_2$ model on the triangular lattice and is relevant for the sodium cobaltate superconductors Na$_x$CoO$_2\cdot y$H$_2$O near $x=1/3$ \cite{Zhou2008,Zheng2006}. The second example is a $d_{x^2-y^2}$ superconductor with coexisting ${\bf Q}=(Q_0,Q_0)$ NCMO on the square lattice which may be realized in a doped $t$-$J_1$-$J_2$-$J_3$ model on the square lattice. Many strongly correlated materials, from high-T$_c$ cuprates to heavy-fermion compounds, exhibits the $d$-wave superconductivity \cite{Tsuei2000,Pfleiderer2009} in proximity to \cite{Lee2006} or coexisting with \cite{Pfleiderer2009,Sigrist1991} magnetic orders. Our findings suggest that Majorana fermions may exist in correlated electron materials with magnetic frustration and nodal superconductivity.

We begin with a general discussion. The low-energy excitations of a nodal superconductor are massless Dirac fermions with linear dispersion.
Our basic idea is to create a topological superconductor
by adding a proper mass to the nodal fermions. Consider the spin-singlet case with $n$ pairs of isolated nodes located at crystal momenta ${\bf \pm q}_\ell$, $\ell=1,\dots n$. Expanding around the nodes, the low-energy BCS Hamiltonian describing the quasiparticle excitations has a generic form in the Nambu basis
$\Psi_{\ell\tk}\equiv(c_{{\bf q}_\ell+\tk,\uparrow},c^\dagger_{-{\bf q}_\ell-\tk,\downarrow};c_{-{\bf q}_\ell+\tk,\downarrow},-c^\dagger_{{\bf q}_\ell-\tk,\uparrow})^T$ for each pair of nodes at opposite momenta:
\begin{equation}\label{h0}
\mathcal{H}_{\rm eff}=\sum_{\ell,\tk}\Psi_{\ell\tk}^\dagger{H}_{\ell\tk} \Psi_{\ell\tk},\quad H_{\ell\tk}=({\bf n}_{\ell\tk}\cdot\vec\tau)\sigma_z,
\end{equation}
and ${\bf n}_{\ell\tk}=(\text{Re}{\Delta_{{\bf q}_\ell+\tk}},-\text{Im}{\Delta_{{\bf q}_\ell+\tk}},\xi_{{\bf q}_\ell+\tk})=\sum_{\alpha}k_\alpha\vec v_\alpha^\ell+ {\cal O}(|\tk|^2)$ with $\alpha=x,y$. Here $\Delta_\tk$ is the pairing gap function and $\xi_\tk$ the kinetic energy. $\vec\tau$ and $\vec\sigma$ are Pauli matrices operating in the particle-hole (Nambu) and spin sectors respectively. A 
unitary rotation $U\equiv\exp[\imth\vec\phi_\ell\cdot\vec\tau]$ turns $H_{\ell\tk}$ into $U^\dagger H_{\ell\tk}U=(k_1\tau_x+k_2\tau_y)\sigma_z$ where $k_{1,2}$ are linearly-independent combinations of $k_x$ and $k_y$ \footnote{This is true as long as $\vec v_x^\ell\times\vec v_y^\ell\neq0$, which is always satisfied by a nodal Dirac point. $(k_1,k_2)$ can be viewed as the two-dimensional momenta in a another coordinate system.}.

Let's focus on the $\ell$-th pair of nodes at $\pm{\bf q}_l$. Clearly, a gap will be generated by adding a generic mass term of the form $\tau_0(\sigma_x\text{Re}M-\sigma_y\text{Im}M)$ to the Hamiltonian,
\begin{equation}
U^\dagger H_{\ell\tk}^\prime U=
(k_1\tau_x+k_2\tau_y)\sigma_z+\tau_0(\sigma_x\text{Re}M-\sigma_y\text{Im}M),
\label{heff}
\end{equation}
where $M$ is a complex order parameter. This is nothing but the effective Hamiltonian for proximity induced $s$-wave superconductivity on the surface of a 3D topological insulator \cite{Fu2008}, with $M$ playing the role of the superconducting (SC) order parameter.
The latter is known to contain a single MBS in the vortex core. In the present context of singlet nodal superconductors, the physical origin of the local order $M$ turns out to be a \emph{non-collinear} (coplanar) magnetic order (NCMO) described by
\begin{eqnarray}\label{magnetic mass}
\mathcal{H}_{\mathrm cp}&=&\sum_{\bf r}\big[M(S_{\bf r}^x+iS_{\bf r}^y)e^{i2{\bf q}_0\cdot{\bf r}}+h.c.\big]
\nonumber \\
&=&\sum_{\ell\tk} Mc^\dagger_{{\bf q}_\ell+\tk+{\bf q}_0,\uparrow}c_{{\bf q}_\ell+\tk-{\bf q}_0,\downarrow}+~h.c.,
\end{eqnarray}
where $S^a_{\bf r}=\sum_{\alpha\beta}c^\dagger_{{\bf r}\alpha}\sigma^a_{\alpha\beta}c_{{\bf r}\beta}/2, a=x,y,z$ are the spin operators at site ${\bf r}$. When the ordering wavevector $2{\bf q}_0$ connects the nodes at $\pm{\bf q}_\ell$, the magnetic scattering generates precisely the mass term in Eq.~(\ref{heff}). Since $SO(3)$ spin-rotational symmetry is completely broken, such a NCMO bears a topologically stable point defect \cite{Kawamura1984} characterized by the nontrivial homotopy $\pi_1\big(SO(3)\big)=\mathbb{Z}_2$. The SC vortex in Fu-Kane model \cite{Fu2008} maps exactly to such a stable point defect of NCMO $M$ in (\ref{heff}). Therefore, \emph{there is a non-Abelian MBS in each stable point defect of non-collinear magnetic order}.

Remarkably, the NCMO gives rise to a non-Abelian topological superconductor since, among the two (even and odd) combinations $c_{\tk,e(o)}\equiv\frac1{\sqrt2}(c_{{\bf q}_\ell+\tk,\uparrow}\pm c_{-{\bf q}_\ell+\tk,\downarrow})$, the odd combination is driven into the topologically nontrivial weak-pairing phase \cite{Read2000} by the mass gap, while the even one to the trivial strong-pairing phase.
The situation is analogous to a doubled-layer $\nu=1/2$ fractional quantum Hall system, where the Abelian (331) state can be driven to a non-Abelian pfaffian state by interlayer tunneling \cite{Ho1995,Read1996,Read2000}.
The existence of a single MBS in the SC vortex core is thus implied by the vortex-boundary correspondence \cite{Read2000}.
Note that the existence of a single MBS will not be affected by the other nodal fermions \cite{Sato2010}. They are spin-flip scattered either to finite energy away from the Fermi level or, in special cases, to different nodes not connected by pairing; they either remain gapless or are gapped out in the magnetic sector.

We stress that it is crucial to require the magnetic order to be \emph{non-collinear}: a collinear spin order, such as
%
$\mathcal{H}_{\mathrm cl}=2m\sum_{\bf r}S_{\bf r}^x\cos(2{\bf q}_0\cdot{\bf r})=\sum_{\ell\tk}\sum_\sigma m c^\dagger_{{\bf q}_\ell+\tk+{\bf q}_0,\sigma}c_{{\bf q}_\ell+\tk-{\bf q}_0,\bar\sigma}+h.c.,$
%
not only drives the Nambu pair $(c_{{\bf q}_0+\tk,\uparrow},c^\dagger_{-{\bf q}_0-\tk,\downarrow})$, but also $(c_{-{\bf q}_0+\tk,\uparrow},c^\dagger_{{\bf q}_0-\tk,\downarrow})$ into the weak-pairing phase, creating two copies of weak-pairing $p+\imth p$ superconductors with two MBSs in the vortex core and two counter-propagating Majorana modes on the edge. Thus, there will be no stable MBSs in this case
since the two branches can scatter and open up a gap in the energy spectrum.

On general grounds, NCMO can be realized in frustrated systems from the residual spin-spin interactions between the nodal fermions in the SC state. Its presence (\ref{magnetic mass}) breaks both inversion and spin rotational symmetry. Thus, the spin-singlet pairing can mix with triplet pairing. When the triplet pairing amplitude is small compared to $|M|$, the system will stay in the gapped non-Abelian topological phase. In the opposite limit, a dominant one-component chiral triplet pairing state is well known to be in the non-Abelian weak-pairing phase \cite{Read2000,Ivanov2001}. Therefore we expect that the non-Abelian topological superconductor to be stable against the mixing between singlet and triplet pairing. We next demonstrate the above predictions with direct calculations in two specific examples on the triangular and the square lattices.

\begin{figure}
 \includegraphics[width=0.22\textwidth]{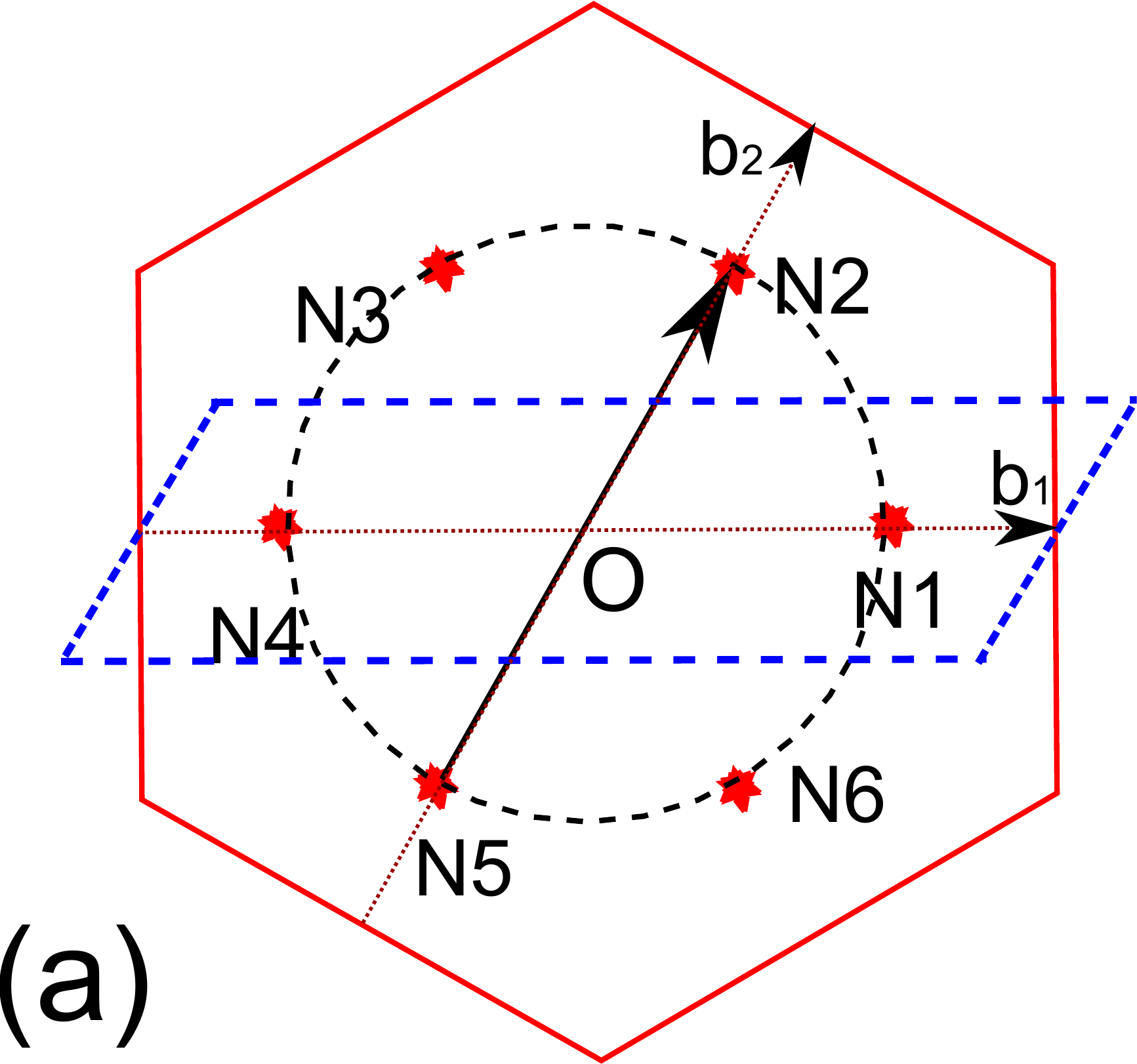}\;\includegraphics[width=0.22\textwidth]{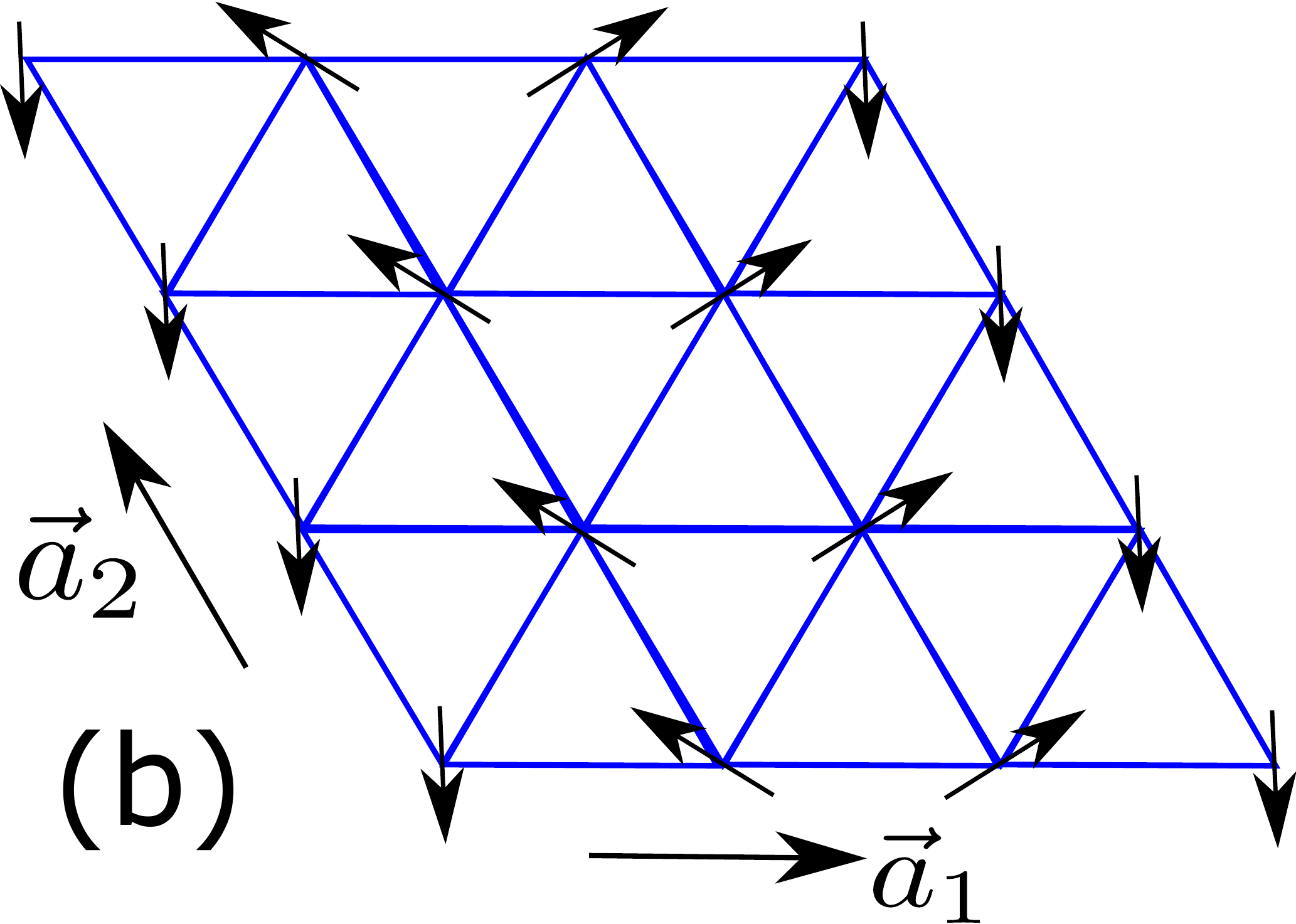}
\caption{(color online) (a) The 1st BZ of the triangular lattice and the $6$ nodes ($N_i$) of the 2nd NN $d+\imth d$ pairing gap function. The normal state FS (dashed circle) crosses the gap nodes at doping $x_c$ for the nodal chiral superconductor. The arrow indicates the wavevector of the $1\times3$ NCMO shown in (b) with the magnetic zone shown by dashed parallelogram.
$\vec a_{1,2}$ are two primitive lattice vectors. The reciprocal vectors are $\vec b_{1,2}$ with $\vec a_i\cdot\vec b_j=\delta_{i,j}$.}\label{fig:tri}
\end{figure}

We start with the nodal chiral superconductors on the triangular lattice proposed for the SC state of hydrated sodium cobaltates \cite{Takada2003}. Recent NMR measurements find strong evidence for singlet pairing \cite{Fujimoto2004,Zheng2006,Zheng2006a} with nodal excitations at a critical doping $x_c\approx0.26$ \cite{Zheng2006a}. Specifically, it was shown \cite{Zhou2008} that 2nd nearest neighbor (NN) $d+\imth d$ pairing can be the dominant pairing channel on the electron doped triangular lattice where the complex gap function has $6$ isolated zeros inside the 1st Brillouin zone (BZ). The Fermi surface (FS) crosses these nodes at a critical doping $x_c$, producing 6 Dirac points as shown in Fig.~\ref{fig:tri}(a). The SC states at $x<x_c$ and $x>x_c$ are separable by a topological phase transition.
We thus consider a simple effective pairing Hamiltonian
\begin{equation}
\mathcal{H}=\sum_{\tk\sigma}\xi_\tk c_{\tk\sigma}^\dagger c_{\tk\sigma}+\sum_\tk (\Delta_\tk c_{\tk\uparrow}c_{-\tk\downarrow}+h.c.),
\label{tri}
\end{equation}
where $\xi_\tk$ is the band dispersion with hopping amplitudes $(t_1,t_2,t_3)=(-202,35,29)$ meV for the first $3$-NN \cite{Zhou2008}. $\Delta_\tk=2\Delta_2\big[\cos(k_1-k_2)+e^{\imth2\pi/3}\cos(2k_1+k_2)+e^{\imth4\pi/3}\cos(k_1+2k_2)\big]$ is the 2nd NN $d+\imth d$ pairing gap function in the basis $\tk=k_1\vec b_1+k_2\vec b_2$ shown in Fig.~\ref{fig:tri}(a). The 6 Dirac nodes ($N_i,~i=1,\cdots,6$) are located at $(k_1,k_2)=\pm(2\pi/3,0)$, $\pm(0,2\pi/3)$, and $\pm(2\pi/3,-2\pi/3)$. Such a $d+\imth d$ superconductor exhibits quantized spin Hall conductance \cite{Read2000} associated with the winding number $W$ of the unit vector $\hat{n}_\tk=(\text{Re}\Delta_\tk,-\text{Im}\Delta_\tk,\xi_\tk)/E_\tk$ where $E_\tk=\sqrt{\xi_\tk^2+|\Delta_\tk|^2}$. When the FS lies inside the Dirac points ($x>x_c$), $W=-2$ and there are two counter-clockwise-propagating chiral fermions on the edge; each is charge neutral but carries spin $\hbar/2$ \cite{Zhou2008}. When the FS encloses the six gap nodes ($x<x_c$), $W=4$ and there are four spin-carrying chiral fermions on the edge.

\begin{figure}
 \includegraphics[width=0.5\textwidth]{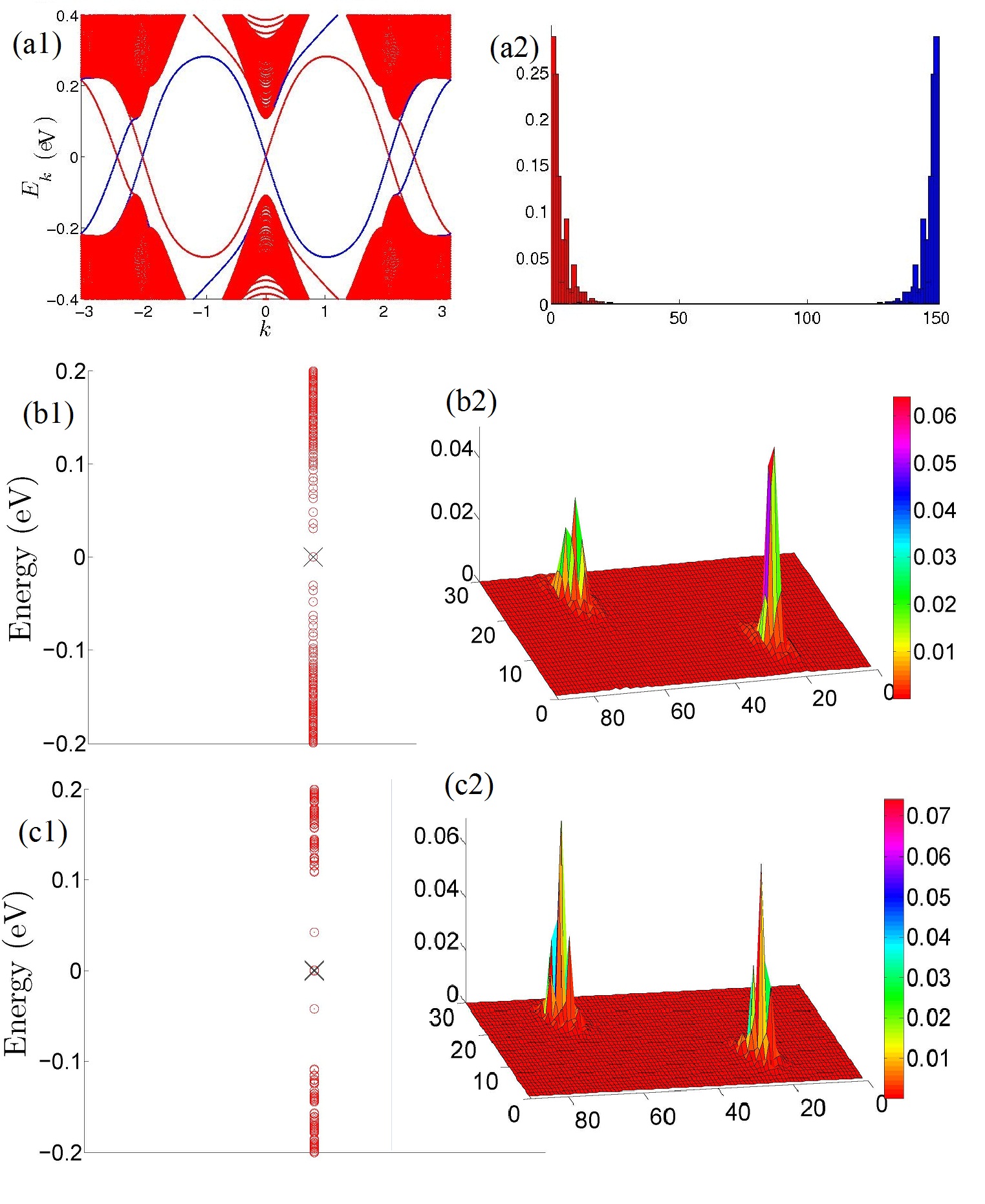}
\caption{(color online) Edge state spectrum (a1) obtained on a cylinder of length $L_1=150$. Vortex spectrum (b1) and magnetic defect spectrum (c1) obtained on $90\times30$ triangular lattices with periodic boundary conditions. The momentum $k$ of the Majorana mode in (a1) is along $\vec b_2$ direction in the magnetic zone in Fig.~\ref{fig:tri}. Right column: the density profiles of the zero-energy MBS. The SC vortex-antivortex pair (b2) and the pair of magnetic defects (c2) are placed at $(r_1,r_2)=(23.5,8.5)$ and $(68.5,23.5)$, where ${\bf r}=r_1\vec a_1+r_2\vec a_2$.}
\label{fig:1x3NC}
\end{figure}

A NCMO described by $\mathcal{H}_{cp}$ in Eq.~(\ref{magnetic mass}), with $2{\bf q}_0$ pointing from $N_{i+3}$ to $N_i$ (Fig.~\ref{fig:tri}a), produces a mass gap for the nodal fermions as discussed above. The magnetic order corresponds to the $1\times3$ coplanar pattern shown in Fig.~\ref{fig:tri}(b). The nodal chiral superconductor is thus turned into a non-Abelian topological superconductor with a winding number $W=+1$. Similar to a spinless $p+\imth p$ superconductor \cite{Read2000}, it supports a single MBS in the vortex core and on the sample edge. To demonstrate the latter, we calculate explicitly the edge spectrum
of $\mathcal{H}+\mathcal{H}_{cp}$ on a cylinder with two parallel edges along the $\vec{a}_2$ direction. The solutions of the BdG equations \cite{supp} are shown in Fig.~\ref{fig:1x3NC}(a) for $\Delta_2=150$ meV and $M=200$ meV. The bulk excitations are completely gapped since the scattering wave vector $2{\bf q}_0$ not only connects the Nambu pair $(N_5,N_2)$ but also $(N_3,N_4)$ and $(N_1,N_6)$ in the magnetic sector.
There is a single branch of gapless Majorana mode crossing $k=0$ that is localized at the edges. We also performed direct calculations of the SC vortex and the magnetic defect spectrum on $90\times30$ periodic lattices in the presence of a vortex-antivortex pair and a pair of stable point defects respectively \cite{supp}. The results are shown in Fig.~\ref{fig:1x3NC}(b) and (c). Clearly, a single zero-energy MBS emerges with a density profile localized in the SC vortex core and at the magnetic defect. Note that since the topological superconductor is in the gapped phase, its stability is protected against perturbations that are not strong enough to destroy the gap and create a quantum phase transition into a different state. As a result, the non-Abelian topological phase supporting MBS proposed here is not limited to very particular parameters and will remain stable when, {\em e.g.} small variations in doping around $x_c$ cause the FS to deviate from the gap nodes, or a small NN pairing component induced by a subdominant NN exchange $J_1$ causes the gap nodes to shift and the magnetic ordering wave vector not to connect precisely the pair of nodes at opposite momenta.
%

To see how the $1\times3$ NCMO can arise microscopically, we consider the 2nd NN antiferromagnetic Heisenberg model, \ie the $J_2$ model, on the triangular lattice. The classical ground state is well-known to have $3\times3$ NCMO \cite{Katsura1986,Jolicoeur1990}: on each of the three sublattices connected by 2nd-NN bonds the spins exhibit 120 degree coplanar order. There is a large ground state degeneracy due to the relative spin orientations.
This $3\times3$ NCMO already induces a single MBS crossing $k=0$ in the edge spectrum \cite{supp}. Quantum fluctuations would lift the degeneracy through the order-due-to-disorder mechanism \cite{Villain1980,Henley1989,Chubukov1992}, and the true quantum ground state has the $1\times3$ order. We have carried out a Schwinger-boson large-$S$ expansion study \cite{supp} of the spin-$S$ Heisenberg $J_2$ model and found that when $S$ is larger than a critical value $S_c\approx0.17$ (such as for spin-$1/2$), the system develops the $1\times3$ NCMO shown in Fig.~\ref{fig:tri}(b). This suggests that if the residual interactions between the nodal fermions are dominated by the 2nd NN Heisenberg exchange $J_2$, the $1\times3$ NCMO is likely to develop with $2{\bf q}_0$ connecting the gap nodes at opposite momenta as shown in Fig.~\ref{fig:tri}. More intriguingly, to the extent that $J_2$ would favor a 2nd NN $d+\imth d$ resonance valence bond pairing state, it is likely that both the nodal chiral superconductor and the NCMO can emerge from the same exchange interaction in a doped $t$-$J_2$ model.

\begin{figure}
 \includegraphics[width=0.22\textwidth]{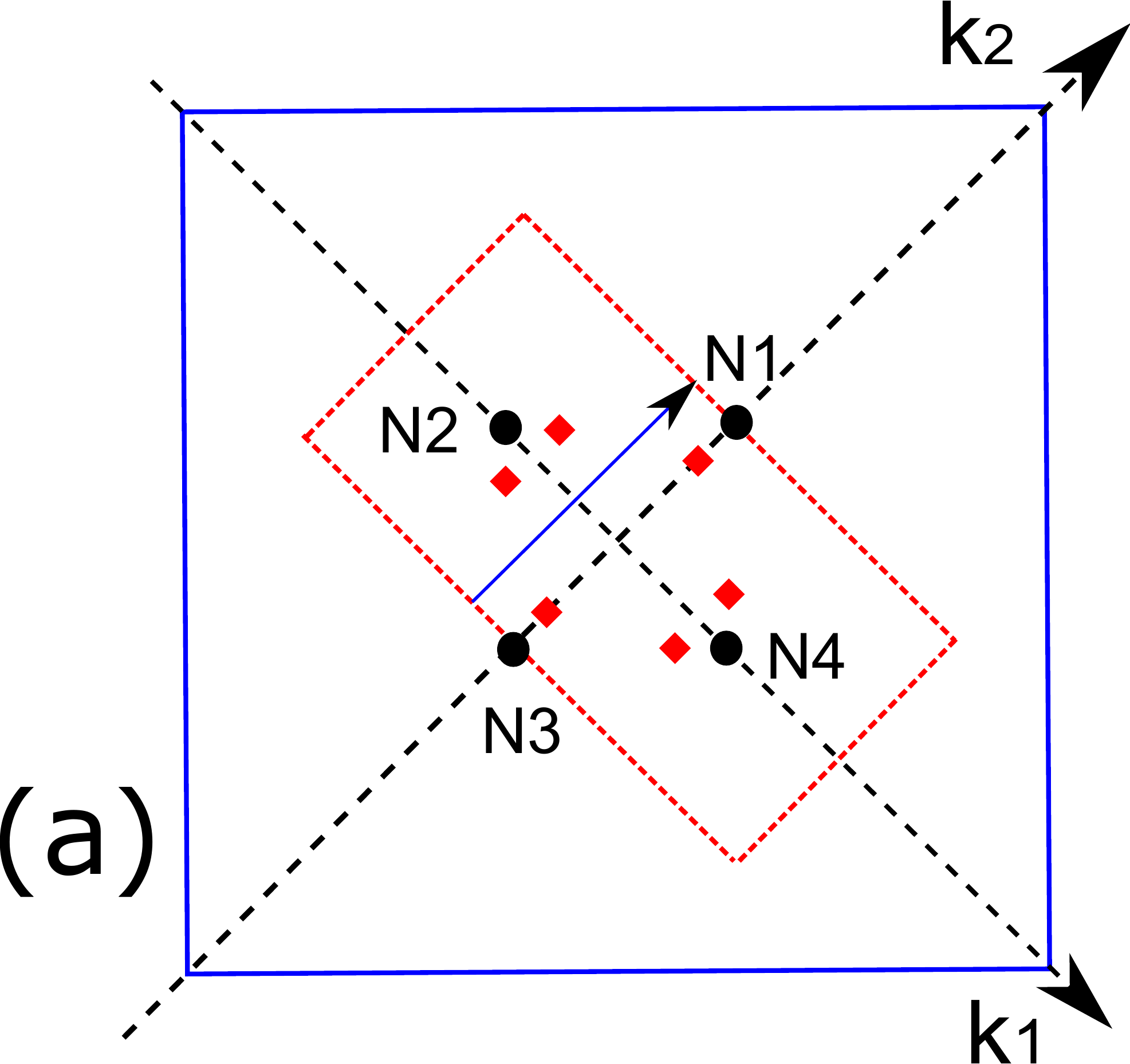}\;\includegraphics[width=0.24\textwidth]{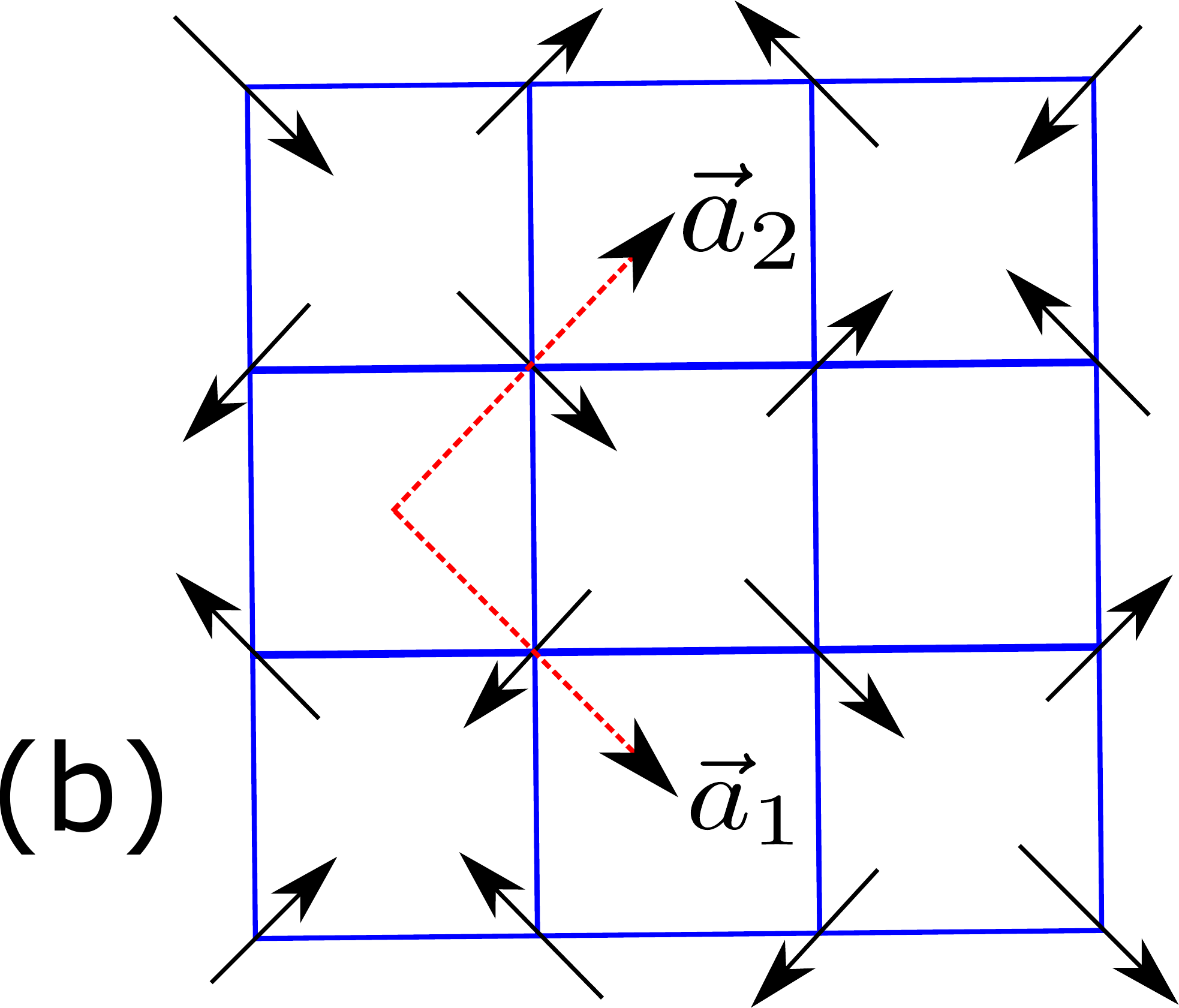}
\caption{(color online) (a) The 1st BZ of the square lattice. Black circles denote the four nodes ($N_i$) where the FS intersects the nodal lines of the $d_{x^2-y^2}$ pairing gap function. The NCMO with ${\bf Q}_0=(\pi/2,\pi/2)$ (black arrow) turns them into 6 red diamonds inside the magnetic zone (red rectangle). (b) The spin configuration of the $(\pi/2,\pi/2)$-NCMO.}
\label{fig:square}
\end{figure}

\begin{figure}
 \includegraphics[width=0.4\textwidth]{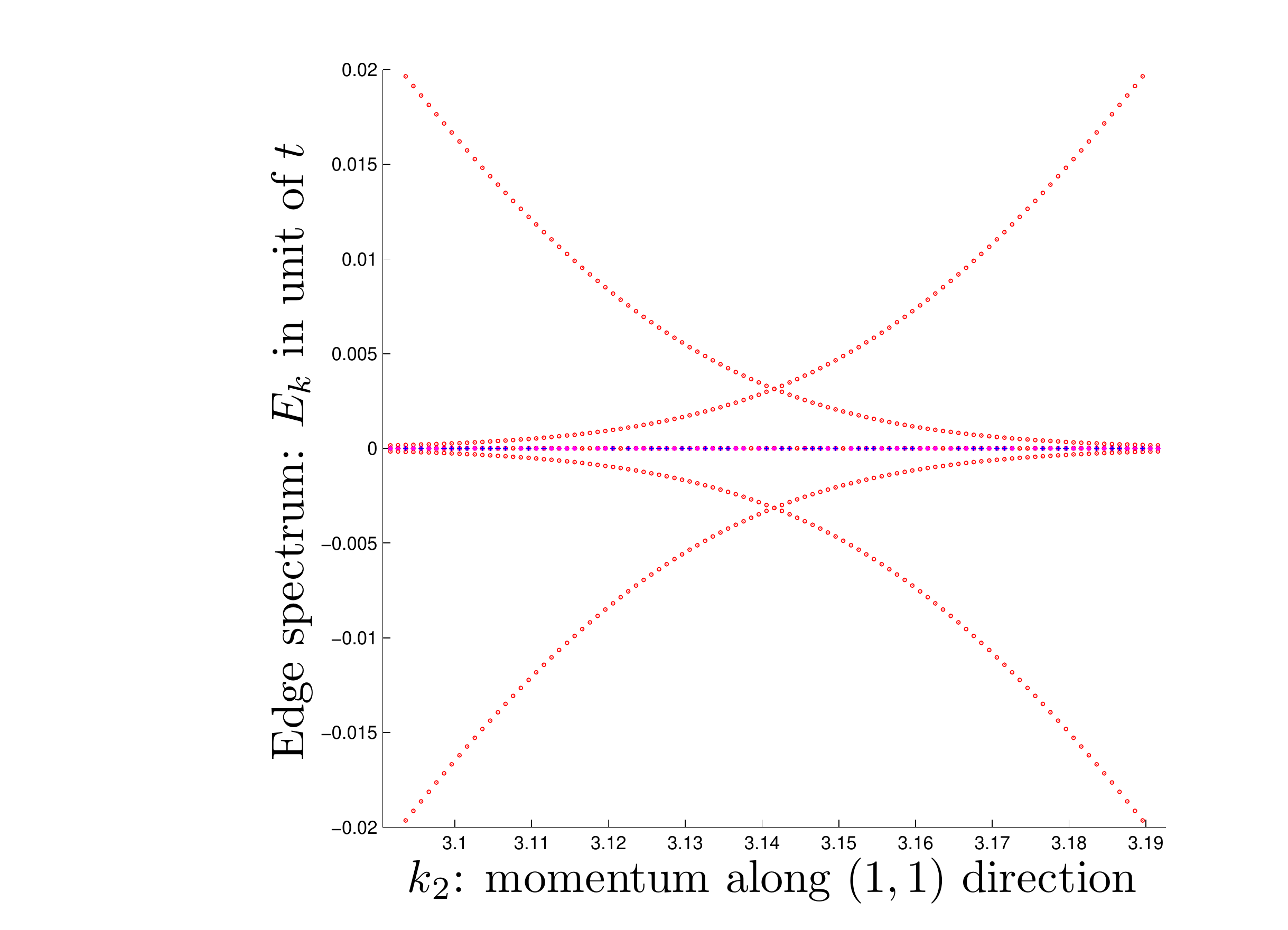}
\caption{(color online) Edge spectra as a function of momentum $k_2$ near the magnetic zone boundary for the $d_{x^2-y^2}$ superconductor with $(\pi/2,\pi/2)$-NCMO. The spectrum is zoomed in around $k_2=\pi$ where a flat band of zero energy MBS is localized on the two edges (blue and magenta). Red dispersing lines are the bulk states with nodes located at $k_2\approx\pi\pm0.04$.}\label{fig:1x4small}
\end{figure}

We turn to the 2nd example of a more familiar nodal $d_{x^2-y^2}$ superconductor
on the square lattice described by the same pairing Hamiltonian $\mathcal{H}$ in Eq.~(\ref{tri}), but with the dispersion $\xi_\tk=-2t\big[\cos(k_1+k_2)+\cos(k_1-k_2)\big]-\mu$ for NN hopping $t$, and the NN $d_{x^2-y^2}$ pairing gap function $\Delta_\tk=2\Delta_1\big[\cos(k_1+k_2)-\cos(k_1-k_2)\big]$. The momentum is defined as $\tk=k_1\vec b_1+k_2\vec k_2$ as shown in Fig.~\ref{fig:square}. The four nodal points are located at
$N_{1,3}:(k_1=0,k_2=\pm q_0)$ and $N_{2,4}:(k_1=\pm q_0,k_2=0)$ with $q_0=\arccos(-\frac\mu{4t})$. A NCMO described by $\mathcal{H}_{cp}$ in Eq.~(\ref{magnetic mass}) with ordering momentum ${\bf Q}_0=(Q_0,Q_0)=2{\bf q}_0=2q_0\vec b_2$ gaps out the $(c_{N_3,\uparrow},c^\dagger_{N_1,\downarrow})$ branch and creates a single MBS. Fig.~\ref{fig:square} shows a specific example with $\mu=-2\sqrt{2}t$ and $q_0=\pi/4$, together with the spin configuration. In this case, the commensurate magnetic order cannot gap out all nodes since the Hamiltonian is still invariant under time reversal followed by a lattice translation \cite{Berg2008}. As shown in Fig.~\ref{fig:square}(a), the NCMO turns the original four spin-degenerate nodes (black circles) into 6 non-degenerate ones (red diamonds). The vanished pair of nodes is gapped out by the magnetic mass (\ref{magnetic mass}) and enters the weak-pairing phase. The calculated edge spectra along (1,1) direction is plotted in Fig.~\ref{fig:1x4small} near the magnetic zone boundary, showing the zero energy MBS localized on the parallel edges. Similar results are obtained at other commensurate values of $q_0$ such as $q_0=\pi/3$. Since the gapless bulk excitations are located at different momenta, the MBS near $k_2=\pi$ is expected to be stable against impurities and the mixing with bulk excitations \cite{Sato2010}. For a generic doping, $q_0$ is incommensurate with the lattice and a corresponding incommensurate NCMO could produce a full gap for bulk excitations.


A remarkable feature seen in Fig.~\ref{fig:1x4small} is that the Majorana mode on the edge is dispersionless, \ie it is localized and does not possess a chirality. This boundary zero-energy flat band, which begins and terminates at the reconstructed nodes of bulk excitations, is a direct consequence of the nontrivial $\mathbb{Z}_2$ winding number (topological index of class D in $d=1$ \cite{Schnyder2008}) of the momentum-space Hamiltonian around the nodes \cite{Wang2012} in FIG. \ref{fig:square}(a). The Majorana flat band is analogous to the Fermi arc on the 2D surface of 3D time-reversal symmetry breaking Weyl semimetals proposed for pyrochlore iridates \cite{Wan2011}. Nevertheless, the time reversal symmetry breaking by the magnetic order (\ref{magnetic mass}) can induce a small imaginary part in $\Delta_1$, which would generate a full gap for bulk excitation and a single MBS on the edge dispersing across $k_2=\pi$ with a well-defined chirality \cite{supp}.

It is possible to realize such a NCMO in the Heisenberg $J_1$-$J_2$-$J_3$ model on the square lattice. The classical ground state has NCMO with momenta ${\bf Q}_0=Q_0\vec b_{1,2}$ where $\cos(Q_0)=-J_1/(2J_2+4J_3)$ for $4J_3+2J_2\geq J_1$ and $J_3\geq J_2/2$ \cite{Rastelli1986,Ferrer1993}. There are numerical evidence that the latter survives in the quantum $S=1/2$ Heisenberg $J_1$-$J_2$-$J_3$ model in a wide parameter range \cite{Sindzingre2009,Sindzingre2010}. We thus expect that such non-Abelian magnetic $d$-wave superconductors 
may be realized in certain parameter regime of the doped $t$-$J_1$-$J_2$-$J_3$ model.

In summary, we proposed a new type of non-Abelian topological superconductors. They emerge when spin-singlet superconductors with isolated nodes coexist with NCMO at the wavevector connecting the nodes at opposite momenta.
Majorana fermions arise in the vortex core and on the edge of such magnetic superconductors. Remarkably, each stable point defect of the non-collinear magnetic order also hosts a single MBS. Since magnetism and unconventional superconductivity are common features of strong correlation, our findings suggest searching for the MBS in correlated materials with magnetic frustration and nodal superconductivity.


We thank S. Zhou for discussions and Aspen Center for Physics for hospitality. This work is supported in part by NSF DMR-0704545 (ZW), DOE DE-FG02-99ER45747 (YML, ZW) and DOE DE-AC02-05CH11231 (YML).

\appendix

\begin{center}
{\Large
Supplementary Materials}
\end{center}

\section{A. Noncollinear magnetic order of $J_2$ Heisenberg model on triangular lattice}

Due to geometric frustration, the triangular lattice antiferromagnet has long been considered as a promising candidate for realizing the spin liquid state \cite{Balents2010}. These spin disordered states are different from conventional symmetry-breaking phases and its low-energy excitations are deconfined fractionalized quasiparticles (such as bosonic/fermionic spinons which carry spin but not electric charge). Strong quantum fluctuations at low temperatures may stabilize such disordered states; whereas in the large-$S$ classical limit, various symmetry breaking magnetic ordered states develop with their low-energy physics dominated by Goldstone modes associated with the broken symmetry, \ie spin waves or magnons. A well-known example is the zero-flux state introduced by Sachdev \cite{Sachdev1992} in the Schwinger-boson representation. As the spin $S$ (or the occupation number of bosonic spinon per site) increases to a critical value \cite{Wang2006}, the spinons at the zone corner condense, leading to the 120 degree ordered state as predicted for the classical Heisenberg $J_1$ model on the triangular lattice. In general a continuous quantum phase transition between a disordered spin liquid and a non-collinear magnetically ordered phase can be described by Bose condensation of spinons \cite{Chubukov1994a}.

Our strategy here is to use the Schwinger-boson representation to study the $Z_2$ spin liquids for the 2nd nearest neighbor (NN) $J_2$ Heisenberg model on the triangular lattice. We then track down the pattern of the noncollinear magnetic order by studying the Bose condensation of spinons in the neighborhood of the spin liquid state using the Schwinger boson mean field theory.

\subsection{Schwinger-boson representation of $Z_2$ spin liquids in triangular lattice $J_2$ model}

In the Schwinger-boson representation the spin operators ${\bf S}_r$ on lattice site $r$ are written in terms of Schwinger bosons $b_{r\alpha}$
\begin{eqnarray}\label{schwinger boson}
{\bf S}_r=\frac12\sum_{\alpha,\beta=\uparrow,\downarrow}
b^\dagger_{r,\alpha}\vec\sigma_{\alpha,\beta}b_{r,\beta}
\end{eqnarray}
where ${\vec \sigma}$ denotes the three Pauli matrices. The spin value $S$ is fixed by the constraint
\begin{eqnarray}\label{constraint}
\sum_{\alpha=\uparrow,\downarrow}b^\dagger_{r,\alpha}b_{r,\alpha}
=\kappa=2S,~~~\forall~r.
\end{eqnarray}
Here $\kappa=2S$ is a parameter measuring the strength of quantum fluctuations in the system and $\kappa\rightarrow+\infty$ corresponds to the classical limit. Consider two sites $i$ and $j$, there are only two bond variables that preserve the spin rotational symmetry
\begin{eqnarray}
&\notag\hat{B}_{ij}=\frac12\sum_{\alpha}b^\dagger_{i,\alpha}b_{j,\alpha}\\
&\notag\hat{A}_{ij}=\frac12\sum_{\alpha\beta}
\epsilon_{\alpha\beta}b_{i,\alpha}b_{j,\beta},
\end{eqnarray}
where $\epsilon_{\alpha\beta}$ is the rank-2 fully anti-symmetric tensor. Thus, a generic Heisenberg Hamiltonian can be written as
\begin{eqnarray}
H=\sum_{i,j}J_{ij}{\bf S}_i\cdot{\bf S}_j=\sum_{i,j}J_{ij}(-\hat{A}_{ij}^\dagger\hat{A}_{ij}
+\hat{B}_{ij}^\dagger\hat{B}_{ij}).
\end{eqnarray}
At the mean-field level, the pairing amplitudes $A_{ij}\equiv\langle\hat{A}_{ij}\rangle=-A_{ji}$ and the hopping amplitudes $B_{ij}\equiv\langle\hat{B}_{ij}\rangle=B_{ji}$ where $A_{ij},B_{ij}$ are complex variational parameters. Enforcing the constraint (\ref{constraint}) globally by a chemical potential $\mu$, we obtain the mean-field Hamiltonian
\begin{eqnarray}
&\label{mfH} H_{MF}=\sum_{i,j}J_{ij}(-{A}_{ij}\hat{A}_{ij}+{B}_{ij}\hat{B}_{ij}+~h.c.)\\
&\notag +\sum_{i,j}J_{ij}(|A_{ij}|^2-|B_{ij}|^2)-
\mu\sum_r(b^\dagger_{r,\alpha}b_{r,\alpha}-\kappa)
\end{eqnarray}

Different types of the spatial ``patterns" of the amplitudes $\{A_{ij},B_{ij}\}$ correspond to different universality classes of the $Z_2$ spin liquids. They are characterized by different symmetry-protected topological orders and classified by the projective symmetry group (PSG) \cite{Wen2002}. In \Ref{Wang2006}, all different $Z_2$ spin liquids preserving the lattice symmetry on the triangular lattice are classified in the Schwinger-boson representation. There are in total 8 different PSGs on an isotropic triangular lattice, corresponding to 8 different universality classes of the $Z_2$ spin liquids.

\begin{figure}
 \includegraphics[width=0.22\textwidth]{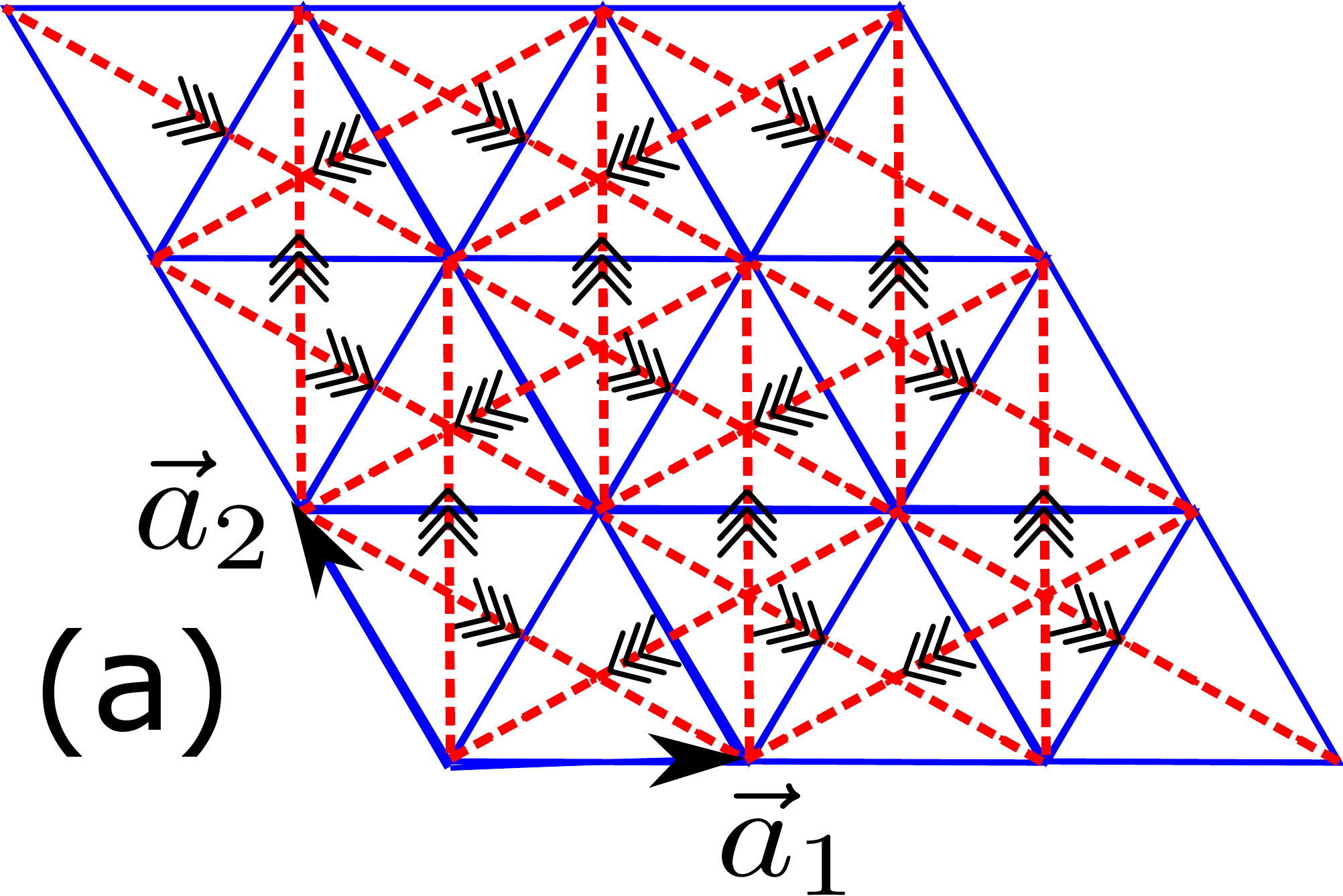}\;\includegraphics[width=0.22\textwidth]{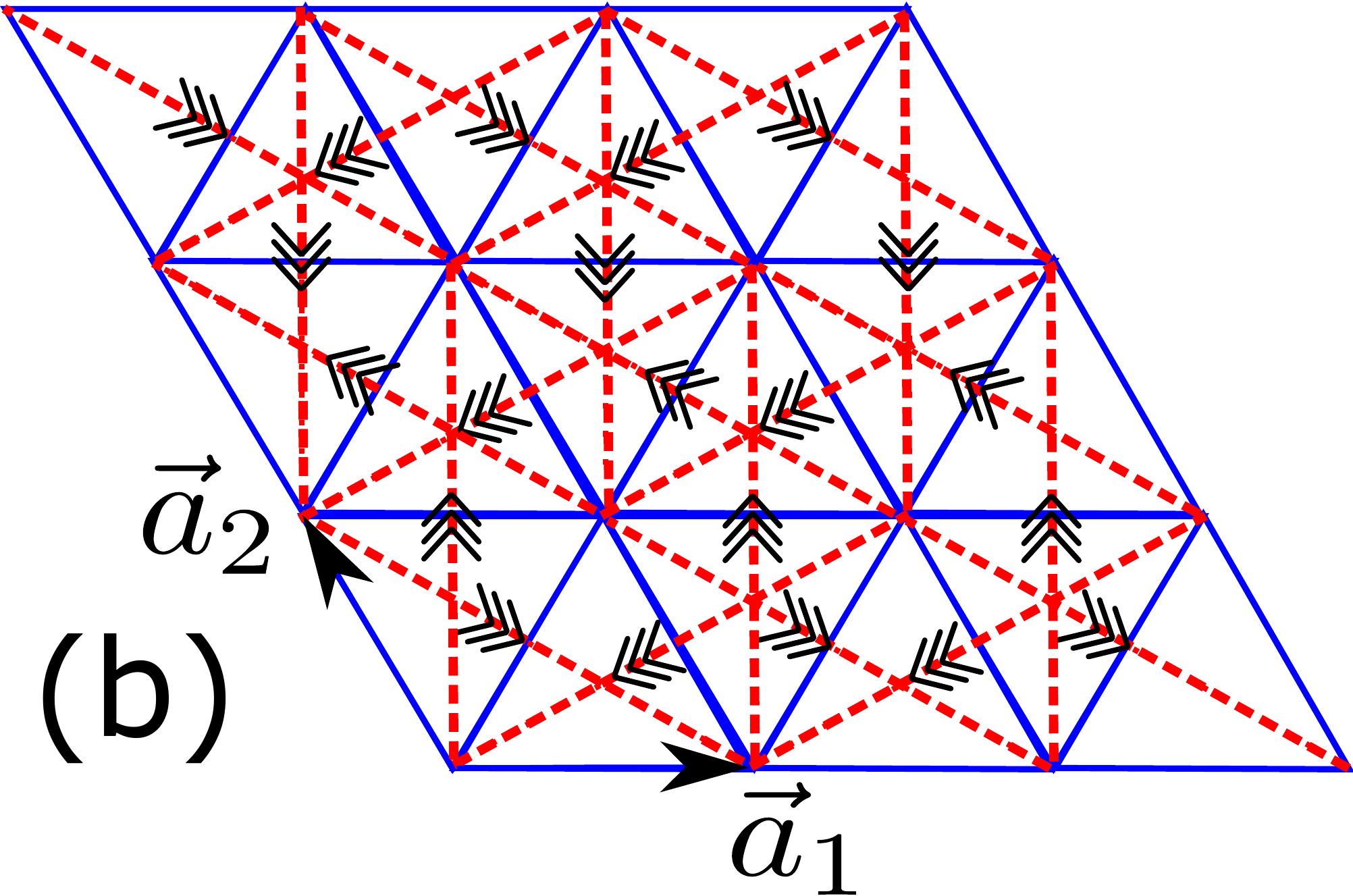}
\caption{(color online) The sign structure of 2nd NN pairing amplitudes $A_{ij}=\frac12\langle b_{i\uparrow}b_{j\downarrow}-b_{i\downarrow}b_{j\uparrow}\rangle$ for the two different $Z_2$ spin liquids: (a) 0-flux-2 state and (b) $\pi$-flux state. Specifically, writing $A_{ij}=\pm\imth A_2$ with $A_2$ a constant, then the plus sign is assigned if $i\rightarrow j$ is along the direction of the triple arrows. $\vec a_{1,2}$ are two primitive lattice vectors. Note that the $\pi$-flux state breaks translation symmetry along the $\vec a_2$ direction and doubles the unit cell.}\label{fig:two ansatz}
\end{figure}

\begin{figure}
 \includegraphics[width=0.5\textwidth]{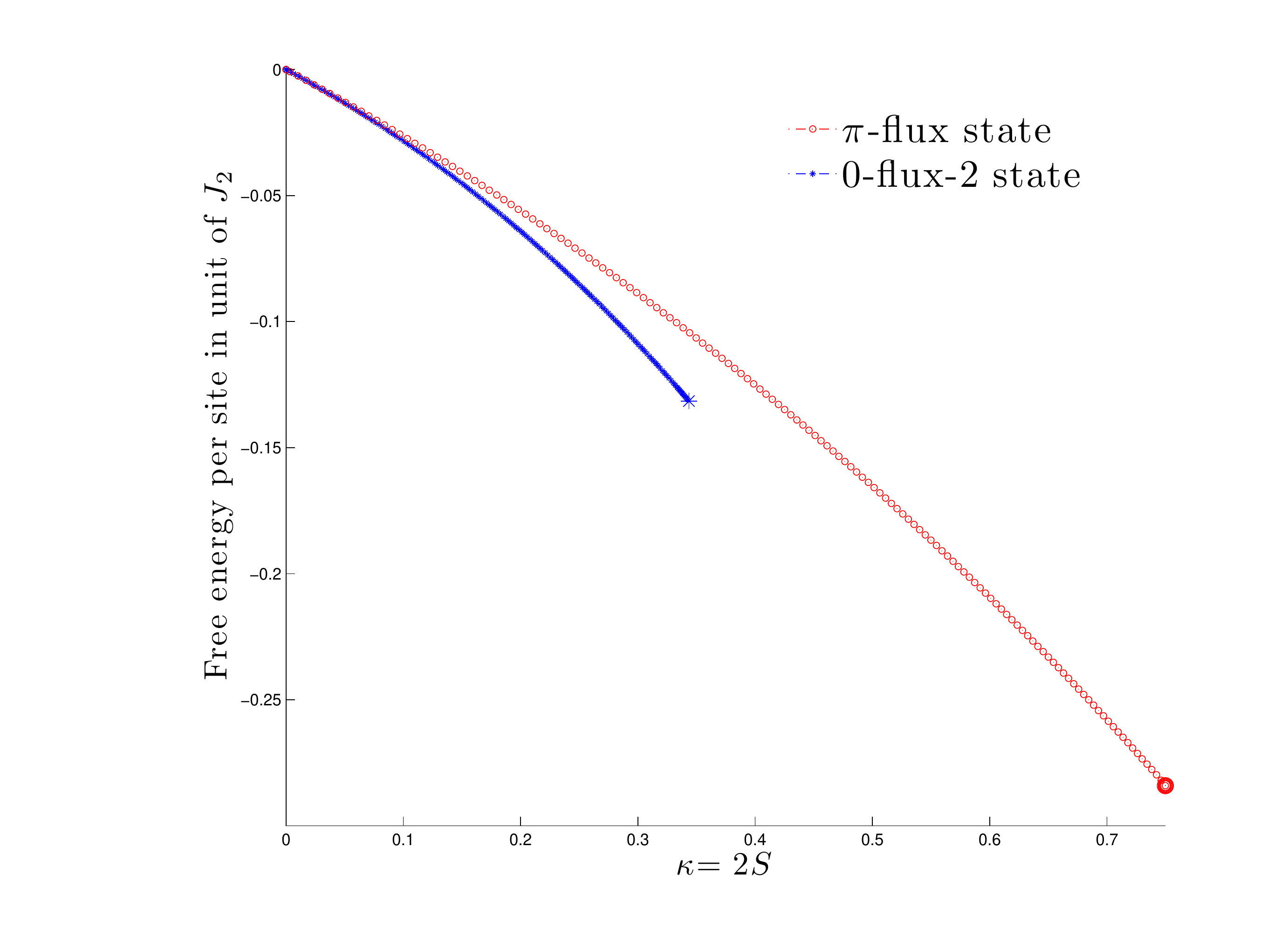}
\caption{(color online) The mean-field ground state energy per site $E_{gs}/(N_sJ_2)=-3|A_2|^2$ of the two $Z_2$ spin liquid states in the $J_2$ Heisenberg model on the triangular lattice. Blue asterisks denote 0-flux-2 state and red circles denote $\pi$-flux state. Apparently 0-flux-2 state has lower energy than its competitor $\pi$-flux state and should be the ground state of $J_2$ model for $\kappa\leq0.34$. When the quantum parameter $\kappa=2S$ increases and becomes larger than critical values $\kappa_c(\text{0-flux-2})\approx0.34$ and $\kappa_c(\pi\text{-flux})\approx0.75$, the spin system goes through a phase transition from spin liquid phases to magnetic ordered phases.}\label{fig:energy}
\end{figure}

Let's focus on those spin liquid states that are possible ground states in the 2nd NN $J_2$ Heisenberg model on an isotropic triangular lattice. We require the 2nd NN mean-field pairing amplitudes to be nonzero: $A_{ij}\neq0$ for $\langle\langle ij\rangle\rangle$. This constraint excludes 6 of the 8 possible $Z_2$ PSGs. Therefore we have only two different $Z_2$ spin liquids with nonzero 2nd NN $A_{ij}$'s: they are labeled as the $\pi$-flux state (following the name given in \Ref{Wang2006}) and 0-flux-2 state (in order to distinguish from the zero-flux state in \Ref{Wang2006}). The 0-flux-2 state corresponds to the $Z_2$ PSG solution $p_1=0,~p_2=p_3=1$ in the notation of \Ref{Wang2006}. Among them the $\pi$-flux state doesn't allow 2nd NN hopping terms, \ie $B_{ij}=0$ for all $\langle\langle ij\rangle\rangle$. The 0-flux-2 state, however, does allow uniform real 2nd NN hopping, \ie $B_{ij}\equiv B_2\in\mathbb{R}$ for all $\langle\langle ij\rangle\rangle$. For both phases, the 2nd NN pairing amplitudes can be made purely imaginary by a gauge choice, with their spatial sign structures shown in Fig.~\ref{fig:two ansatz}.
We carry out self-consistent calculations of the mean-field Hamiltonian (\ref{mfH}) for both the $\pi$-flux and the 0-flux-2 states. For simplicity we only include the 2nd NN pairing amplitudes for the 0-flux-2 state. Thus, strictly speaking the obtained ground state energy for the 0-flux-2 state gives an upper bound of the actual energy. The comparison of the ground state energy (or zero-temperature free energy) for the two states are shown in Fig.~\ref{fig:energy}, which shows that the 0-flux-2 state has a lower energy than the $\pi$-flux state. A nonzero 2nd NN hopping amplitudes $B_2$ would lower the energy of the 0-flux-2 state even further.
Thus the lowest energy $Z_2$ spin liquid state in the triangular lattice $J_2$ model should be the 0-flux-2 state which, as we will show below, is stable against spin order when the quantum parameter is smaller than $\kappa_c\approx0.34$ (see Fig.~\ref{fig:energy}).

\subsection{Spinon condensation and the magnetic order in triangular lattice $J_2$ Heisenberg model}

\begin{figure}
 \includegraphics[width=0.4\textwidth]{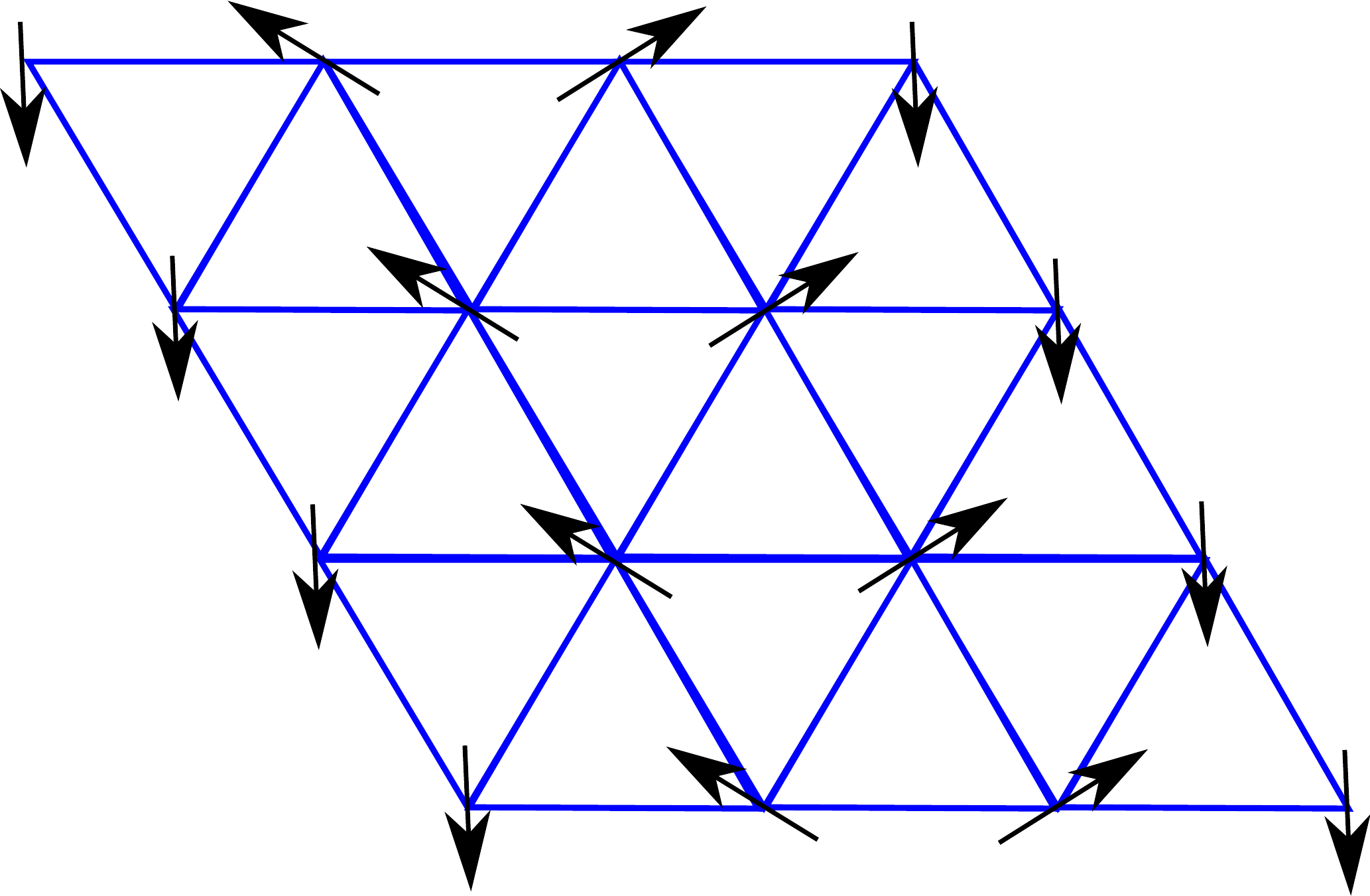}
\caption{(color online) The spin configuration of the $1\times3$ coplanar magnetic order in a $S=1/2$ Heisenberg $J_2$ model on the triangular lattice, obtained from spinon condensation in the 0-flux-2 state.}\label{fig:1x3}
\end{figure}

Now we show how to extract the magnetic ordering pattern in the 0-flux-2 state when $\kappa$ becomes large. We choose $\vec b_{1,2}$ to be the reciprocal lattice vector which satisfies $\vec a_i\cdot\vec b_j=\delta_{i,j},~~i,j=1,2$ where $\vec a_{1,2}$ are the primitive vectors shown in Fig.~\ref{fig:two ansatz}. Expressing the momentum as $\tk=k_1\vec b_1+k_2\vec b_2$, the mean-field Hamiltonian can be written as
\begin{eqnarray}
&\notag H_{MF}=\sum_\tk\begin{bmatrix}b^\dagger_{\tk,\uparrow}\\b_{-\tk,\downarrow}\end{bmatrix}^T\begin{pmatrix}-\mu&J_2A_2g^{(2)}_\tk\\J_2A_2g^{(2)}_\tk&-\mu\end{pmatrix}\begin{bmatrix}b_{\tk,\uparrow}\\b^\dagger_{-\tk,\downarrow}\end{bmatrix}\\
&+N_s\Big[\mu(\kappa+1)+3J_2|A_2|^2\Big],
\label{spinonmf}
\end{eqnarray}
where $N_s$ is the number of lattice sites and
\begin{eqnarray}
g_\tk^{(2)}=-\sin(k_1-k_2)-\sin(k_1+2k_2)+\sin(2k_1+k_2)\notag
\end{eqnarray}
is the 2nd NN pairing structure factor corresponding to Fig.~\ref{fig:two ansatz}(a). The spinon dispersion, obtained from Eq.~(\ref{spinonmf}),
\begin{eqnarray}
\lambda_\tk=\sqrt{\mu^2-|J_2A_2g_\tk^{(2)}|^2}
\end{eqnarray}
has 6 minima at $(k_1,k_2)=\pm(2\pi/3,0)$,~$\pm(0,2\pi/3)$ and $\pm(2\pi/3,-2\pi/3)$ which coincide with the 6 gap nodes $\{N_i\}$ of the 2nd NN $d+\imth d$ pairing gap function\footnote{We verified that 2nd NN hopping amplitudes will not alter the spinon band minima and thus the magnetic ordering pattern upon spinon condensation.}. We denote the momenta at points $N_i$ as $\tk_i$. As the quantum parameter $\kappa$ increases, the chemical potential $\mu$ decreases and the minima of the spinon bands $\min\{\lambda_\tk\}=\sqrt{\mu^2-27|J_2A_2|^2/4}$ approach zero from the positive side. When $\mu=3\sqrt3J_2A_2/2$ the band minima touch zero, the spinons at the 6 momentum points condense into
\begin{eqnarray}
\begin{bmatrix}\langle b_{\tk_i,\uparrow}\rangle\\ \langle b^\dagger_{-\tk_i,\downarrow}\rangle\end{bmatrix}=c_i\begin{bmatrix}1\\(-1)^i\end{bmatrix},~~~i=1,2,3,4,5,6,
\end{eqnarray}
where $c_i$ are complex numbers. The spatial configuration of the spinons is then given by
\begin{eqnarray}
&\notag\Psi_{\bf r}\equiv\begin{bmatrix}\langle b_{{\bf r},\uparrow}\rangle\\ \langle b_{{\bf r},\downarrow}\rangle\end{bmatrix}=\sum_{i=1}^6e^{\imth\tk_i\cdot{\bf r}}\begin{bmatrix}\langle b_{\tk_i,\uparrow}\rangle\\ \langle b_{\tk_i,\downarrow}\rangle\end{bmatrix}\\
&=\phi_1U_{1}\begin{bmatrix}\omega^x\\ \omega^{-x}\end{bmatrix}+\phi_2U_{2}\begin{bmatrix}\omega^y\\ \omega^{-y}\end{bmatrix}+\phi_3U_{3}\begin{bmatrix}\omega^{y-x}\\ \omega^{x-y}\end{bmatrix}\label{BEC:triangular}
\end{eqnarray}
where ${\bf r}=x\vec a_1+y\vec a_2$ is the position vector of a site and $\omega=\exp(\imth2\pi/3)$. In the above equation, $\phi_{1,2,3}$ are three positive constants and $U_{1,2,3}$ are three $SU(2)$ matrices with
\begin{eqnarray}
\phi_i U_i=\begin{pmatrix}c_i&c_{i+3}\\-c^\ast_{i+3}&c^\ast_i\end{pmatrix},~~~i=1,2,3.\notag
\end{eqnarray}
The condensed spinons give rise to magnetic order and the corresponding spin configuration is
\begin{eqnarray}\label{order:triangularJ2}
{\bf S}_{\bf r}&=&\frac12\Psi_{\bf r}^\dagger\vec\sigma\Psi_{\bf r}=\frac12\text{Tr}\big[\Psi_{\bf r}\Psi_{\bf r}^\dagger\vec\sigma\big]\\
&=&\notag\frac{\phi_1^2}2\text{Tr}\Big\{\begin{bmatrix}1&\omega^{2x}\\ \omega^{-2x}&1\end{bmatrix}U_1^\dagger\vec\sigma U_1\Big\}\\
&+&\notag\frac{\phi_2^2}2\text{Tr}\Big\{\begin{bmatrix}1&\omega^{2y}\\ \omega^{-2y}&1\end{bmatrix}U_2^\dagger\vec\sigma U_2\Big\}\\
&\notag+&\frac{\phi_3^2}2\text{Tr}\Big\{\begin{bmatrix}1&\omega^{2(y-x)}\\ \omega^{2(x-y)}&1\end{bmatrix}U_3^\dagger\vec\sigma U_3\Big\}.
\end{eqnarray}
In general, when all three $\phi_i$'s are nonzero, the magnetic order has the  $3\times3$ spatial pattern on the triangular lattice. More precisely, there are three sublattices connected by the 2nd-NN $J_2$ bonds: on every sublattice the spins form coplanar 120 degree order as in the NN $J_1$ Heisenberg model. When only one $\phi_i$ is nonzero, the corresponding spin configuration is the stripe-like $1\times3$ coplanar ordered shown in Fig.~\ref{fig:1x3}.

The classical ground state of the $J_1$-$J_2$ Heisenberg model with $\alpha=J_2/J_1>1$ is known \cite{Katsura1986,Jolicoeur1990} to have non-collinear magnetic order on the triangular lattice with the following momenta
\begin{eqnarray}
\notag&\tk=(k_1,k_2)=\pm(Q_0,0),~\pm(0,Q_0)~\text{and}~\pm(Q_0,-Q_0),\\
&Q_0=\arccos(-\frac12-\frac1{2\alpha}).\label{vector:triangular}
\end{eqnarray}
In the limit $\alpha\rightarrow+\infty$, they exactly correspond to the magnetic order (\ref{order:triangularJ2}) obtained by condensing spinons in the 0-flux-2 spin liquid state. These classical solutions have a large ground state degeneracy, due to the relative spin orientations between different 2nd-NN sublattices as mentioned earlier. When quantum fluctuations are considered, the degeneracy at the classical level will be lifted through the ``order due to disorder" mechanism \cite{Villain1980,Henley1989}. Therefore the ground state of the quantum Heisenberg $J_2$ model is expected \cite{Chubukov1992} to exhibit magnetic order with one ordering wavevector among those in (\ref{vector:triangular}). In other words, only one of the $\{c_i,~i=1,\cdots,6\}$ in (\ref{BEC:triangular}) is nonzero and the corresponding spin configuration (\ref{order:triangularJ2}) is the $1\times3$ noncollinear order shown in Fig.~\ref{fig:1x3}.

\section{B. Majorana bound states in the triangular lattice model}

All the energy spectra in this work are obtained by solving Bogoliubov-de-Gennes (BdG) equations, i.e. diagonalizing the quadratic Hamiltonians in Eqs.~(5) and (3), on triangular or square lattices with different boundary conditions. For the triangular lattice, the edge spectra are obtained on a cylinder of width $L_1=3\times50=150$, with open boundary condition along the $\vec{a}_1$ direction and periodic boundary condition along the $\vec{a}_2$ direction defined in Fig.~1(b). The edge is chosen to be along the $\vec{a}_2$ direction of the triangular lattice. Hence the momentum $k$ along $\vec{b}_2$ is a good quantum number used to label the edge spectra. On the other hand, the energy spectra with a pair of superconducting (SC) vortices, or a pair of point defects of the non-collinear magnetic order, are obtained on a $90\times30$ torus with periodic boundary conditions in both directions.

To be specific, the BdG Hamiltonian from Eqs.~(5) and (3) can be written as
\begin{eqnarray}
H_{BdG}=\sum_{{\bf r},{\bf r}^\prime}\begin{bmatrix}c^\dagger_{{\bf r},\uparrow}\\c^\dagger_{{\bf r},\downarrow}\\ c_{{\bf r},\downarrow}\\-c_{{\bf r},\uparrow}\end{bmatrix}^T{\bf H}({\bf r}|{\bf r}^\prime)\begin{bmatrix}c_{{\bf r}^\prime,\uparrow}\\c_{{\bf r}^\prime,\downarrow}\\ c^\dagger_{{\bf r}^\prime,\downarrow}\\-c^\dagger_{{\bf r}^\prime,\uparrow}\end{bmatrix}
\end{eqnarray}
where ${\bf r}=x\vec{a}_1+y\vec{a}_2$ labels the lattice site position. The BdG matrix is given by
\begin{eqnarray}
&\notag{\bf H}({\bf r}|{\bf r}^\prime)=\begin{bmatrix}t({\bf r}|{\bf r}^\prime)&\Delta({\bf r}|{\bf r}^\prime)\\ \Delta^\ast({\bf r}|{\bf r}^\prime)&-t^\ast({\bf r}|{\bf r}^\prime)\end{bmatrix}\otimes\hat{I}_{2\times2}+\\
&\frac{\delta_{{\bf r},{\bf r}^\prime}}2\cdot\Big[-\mu\sigma_z\otimes \hat{I}_{2\times2}+\hat{I}_{2\times2}\otimes\big(\vec{M}({\bf r})\cdot\vec{\sigma}\big)\Big],\notag
\end{eqnarray}
where $\mu$ is the chemical potential, $\vec{M}({\bf r})$ is the local magnetic order parameter, $t({\bf r}|{\bf r}^\prime)$ and $\Delta({\bf r}|{\bf r}^\prime)$ represent respectively the hopping and spin-singlet pairing between sites ${\bf r}$ and ${\bf r}^\prime$, and $\hat{I}_{2\times2}$ denotes the $2\times2$ identity matrix and $\sigma_{x,y,z}$ are the three Pauli matrices. For the cobaltate band structure, the electron hopping terms are nonzero for up to 3rd nearest neighbors (NNs): $(t_1,t_2,t_3)=(-202,35,29)~meV$. The nodal SC state corresponds to taking $\mu=-3t_2$ where the normal state Fermi surface crosses the six nodes in the gap function of the 2nd NN $d+\imth d$ pairing. Diagonalizing the matrix ${\bf H}({\bf r}|{\bf r}^\prime)$ yields the energy spectrum.

\begin{figure}
\includegraphics[width=0.47\textwidth]{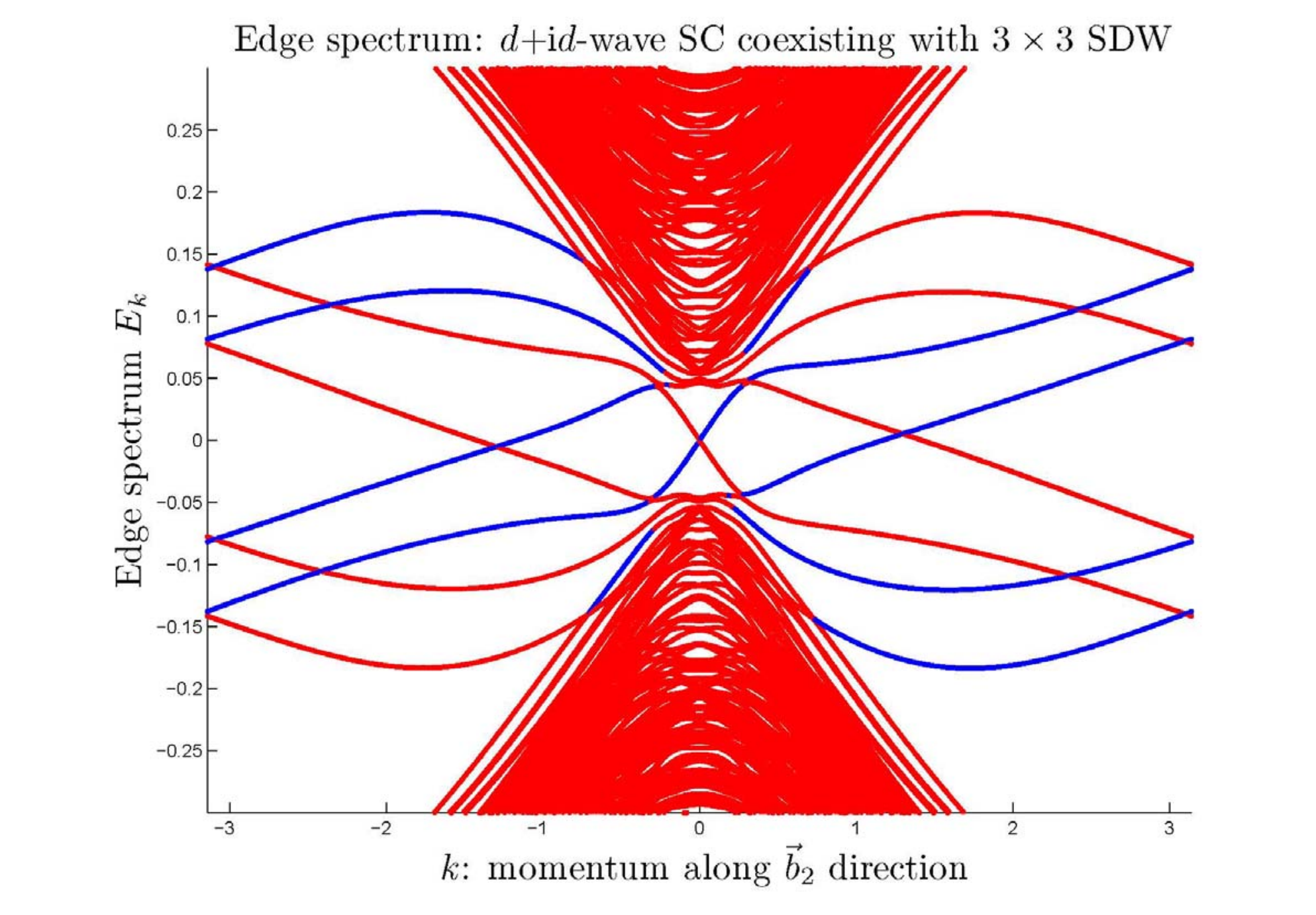}
\caption{(color online) Edge spectrum of 2nd-NN $d+\imth d$ superconductor coexisting with an arbitrary $3\times3$ non-collinear magnetic order (\ref{order:triangularJ2}). Notations are the same as in Fig.~2 of the main text. Among the in-gap edge states separated from the bulk continuum, blue lines represent edge states on one edge, while red lines represent edge states on the other edge. The edge is along (1,1) direction. There is one single clockwise-propagating MBS crossing $k=0$ in the magnetic zone. The data are obtained for $\Delta_2=150~meV$ and $M=100~meV$ for all three ordering vectors.}\label{fig:3x3}
\end{figure}

\subsection{Majorana edge mode with $3\times3$ noncollinear magnetic order}

It turns out that even for an arbitrary classical ground state (\ref{order:triangularJ2}) with $3\times3$ non-collinear magnetic order, when it coexists with the 2nd NN $d+\imth d$ superconductivity on the triangular lattice, there will still be a single Majorana bound state on the edge of the sample. An explicit example of the edge spectrum is shown in Fig.~\ref{fig:3x3}, where a single Majorana mode is localized on each of the two parallel edges and disperses across $k=0$.


\subsection{Absence of Majorana bound states with collinear magnetic order}

\begin{figure}
\includegraphics[width=0.5\textwidth]{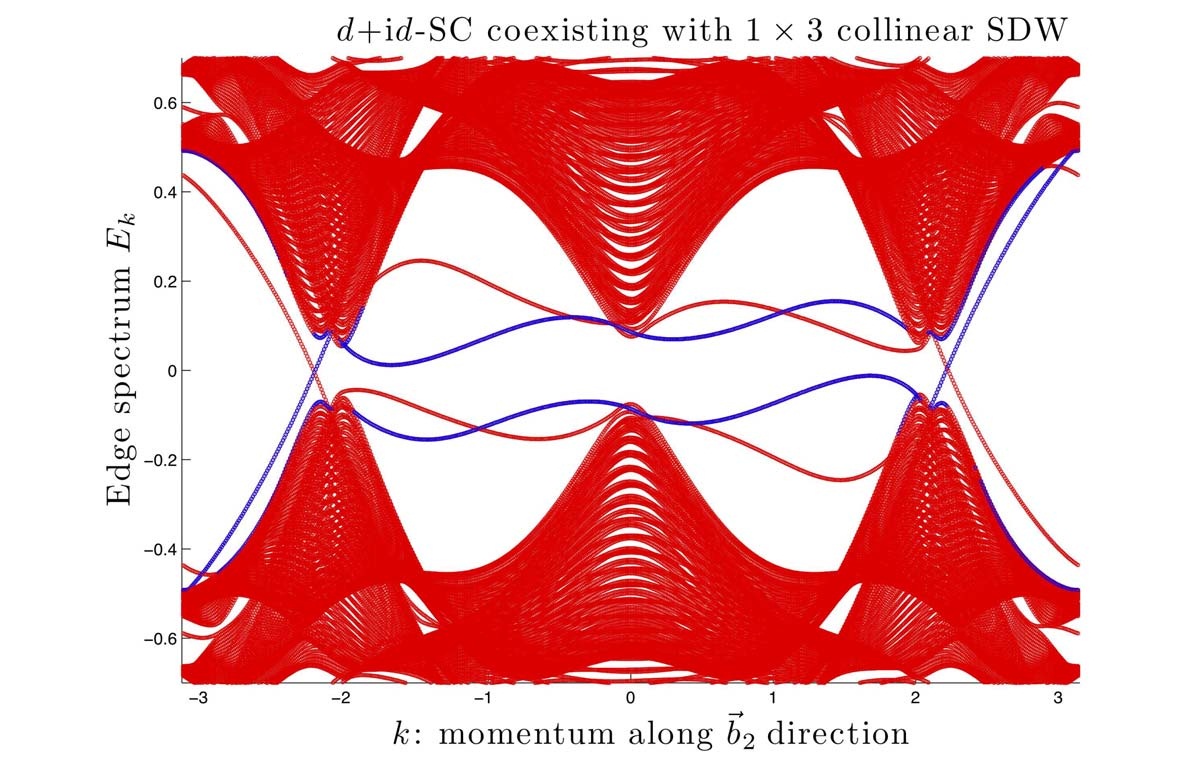}
\caption{(color online) The edge spectra with momentum $k$ along the $\vec b_2$ direction of a 2nd-NN $d+\imth d$ superconductor coexisting with $1\times3$ \emph{collinear} magnetic order with magnetization along the $\hat z$ direction on the triangular lattice. Notations are the same as in Fig.~2 of the main text. Note the absence of Majorana bound state in the spectrum. The data are obtained for pairing energy $\Delta_2=150~meV$ and collinear magnetic mass $m=100~meV$.}\label{fig:1x3C}
\end{figure}

As discussed in the main text, a collinear magnetic order described by $\mathcal{H}_{cl}$ in Eq.~(4) in the main text will not create a Majorana bound state in the vortex core or on the sample edge, although the bulk excitations are fully gapped. In Fig.~\ref{fig:1x3C}, we show explicitly the absence of Majorana mode in the edge spectrum obtained for a 2nd-NN $d+\imth d$ superconductor coexisting with $1\times3$ \emph{collinear} magnetic order with magnetization along the $\hat z$ direction. As before, the edge spectrum is calculated on a cylinder of size $L_1=3\times50=150$.

\subsection{Majorana bound states in the superconducting vortex core}

The single Majorana edge mode implies the existence of a single MBS in the superconducting vortex core, as a manifestation of edge-vortex correspondence \cite{Read2000}. In this subsection, we present an explicit calculation which confirms this correspondence principle even in the presence of the coexisting $1\times3$ noncollinear magnetic order.

\begin{figure}
\includegraphics[width=0.24\textwidth]{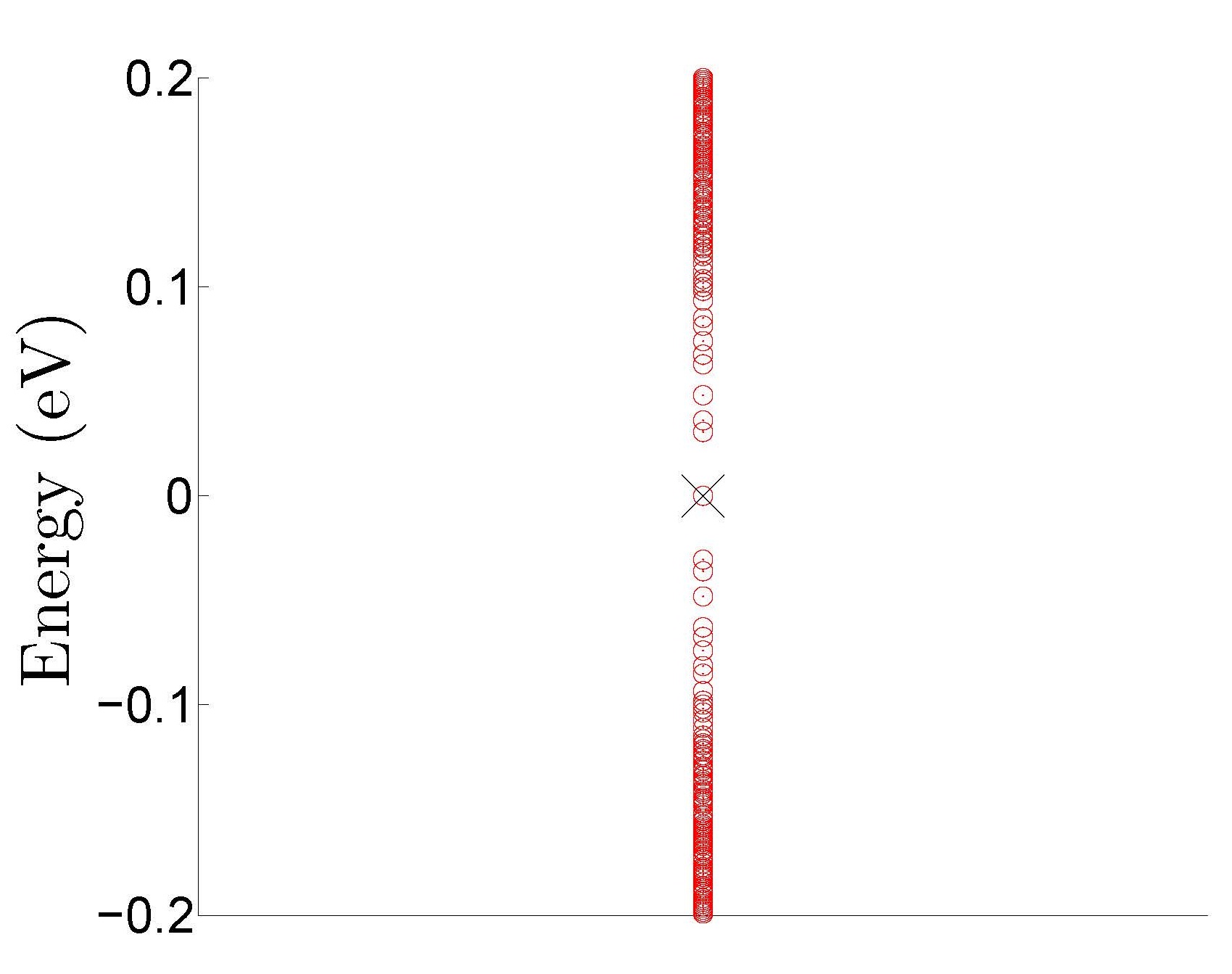}\;\includegraphics[width=0.24\textwidth]{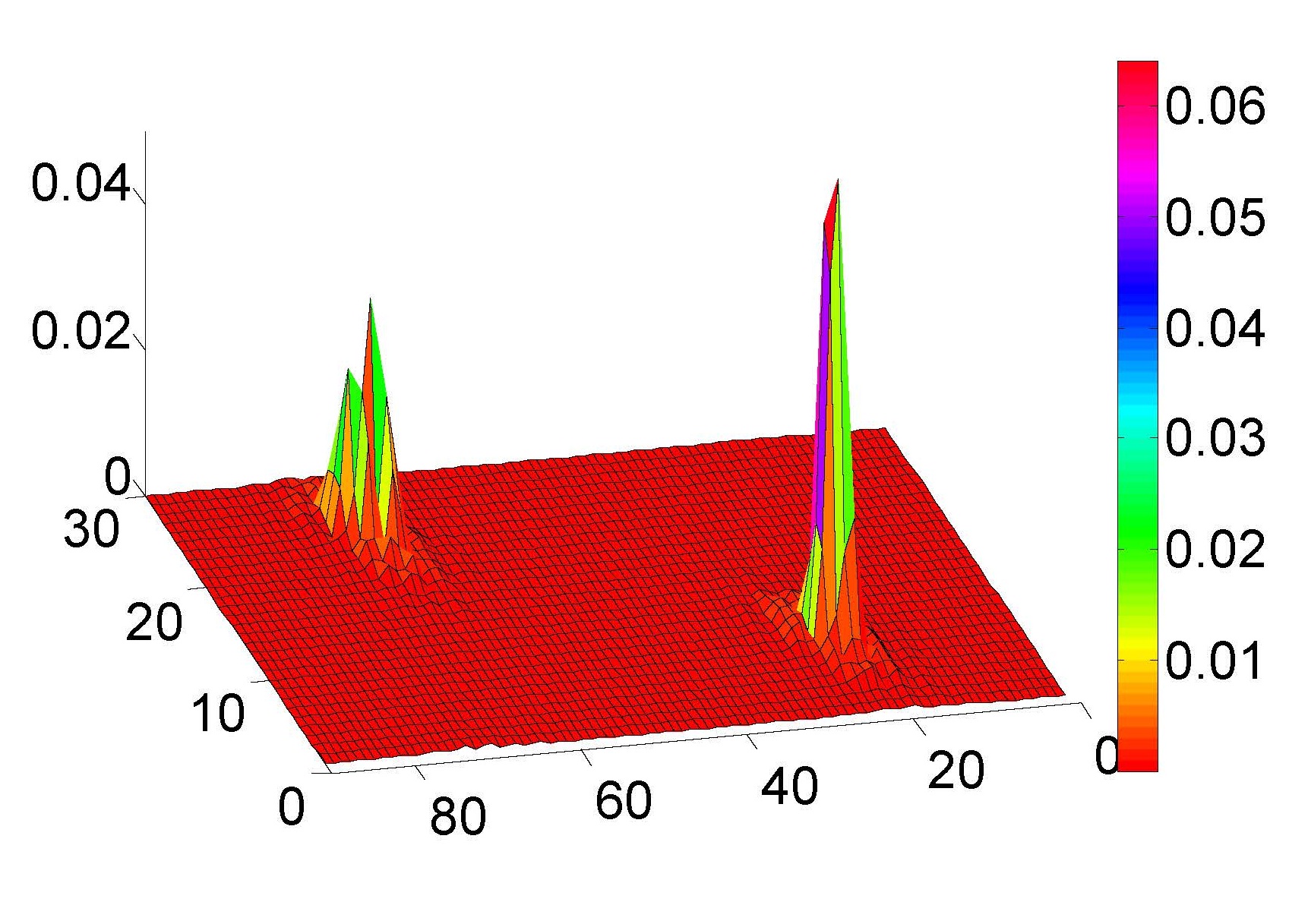}
\caption{(color online) SC vortex spectrum and MBS density profile on the triangular lattice. A vortex-antivortex pair in the SC order parameter are located at $(r_1,r_2)=(23.497,8.497)$ and $(68.497,23.497)$ on a $90\times30$ torus respectively. The zero-energy ($E_0\approx0.016~meV$) MBS (black cross) is separated from the first excited state by an energy $|E_1-E_0|\approx30~meV$. The electron density of the MBS is concentrated in the cores of the SC vortex/antivortex. The bulk energy gap is around $100~meV$. The data are obtained for $\Delta_2=150~meV$ and $M=200~meV$ for $1\times3$ noncollinear magnetic order.}\label{fig:vortex}
\end{figure}

The vortex-antivortex configuration is enforced by assigning the following phase factor to the pairing amplitude between site $(x_1,y_1)$ and $(x_2,y_2)$:
\begin{eqnarray}
&\notag\Delta(x_1,y_1|x_2,y_2)=|\Delta_2|\cdot\\
&e^{2\imth\text{arg}\big(x_2-x_1+\imth(y_2-y_1)\big)}e^{\imth\phi(\frac{x_1+x_2}2+\imth\frac{y_1+y_2}2)}.
\end{eqnarray}
The first phase factor corresponds to the $d+\imth d$-wave pairing, while the second phase factor generates the vortex-antivortex configuration in the SC order parameter. The latter, $e^{\imth\phi}$, is a smooth function defined on a $L_1\times L_2$ torus as
\begin{eqnarray}
&\notag\phi(z=x+\imth y)=\arg{\Big(\vartheta_1(\frac{\pi(z-w_1)}{L_1}|\imth\frac{L_2}{L_1})\bar{\vartheta_1}(\frac{\pi(z-w_2)}{L_1}|\imth\frac{L_2}{L_1})\Big)}\\
&-\frac{2\pi\text{Im}(z)}{L_1L_2}\text{Re}(w_1-w_2), \label{vortex pair angle}
\end{eqnarray}
where the Jacobi theta-function of the 1st type is defined as
\begin{eqnarray}
&\notag\vartheta_1(z|\tau)=\vartheta_1(z,q=e^{\imth\pi\tau})\equiv\\
&\sum_{n=-\infty}^{+\infty}(-1)^{n-1/2}q^{(n+1/2)^2}e^{(2n+1)\imth z},
\end{eqnarray}
with $\text{Im}\tau>0$. Here $\bar{\vartheta_1}$ denotes the complex conjugate of $\vartheta_1$.

Solving the BdG equations in the presence of the SC vortex-antivortex pair, we obtain the vortex energy spectrum which is shown in Fig.~2(b) in the main text and reproduced here in Fig.~\ref{fig:vortex}. Note that there is a zero-energy MBS within the gap, whose electron density is concentrated around the vortex and antivortex.

\subsection{Majorana bound states in the stable point defect of $1\times3$ noncollinear magnetic order}

\begin{figure}
\includegraphics[width=0.24\textwidth]{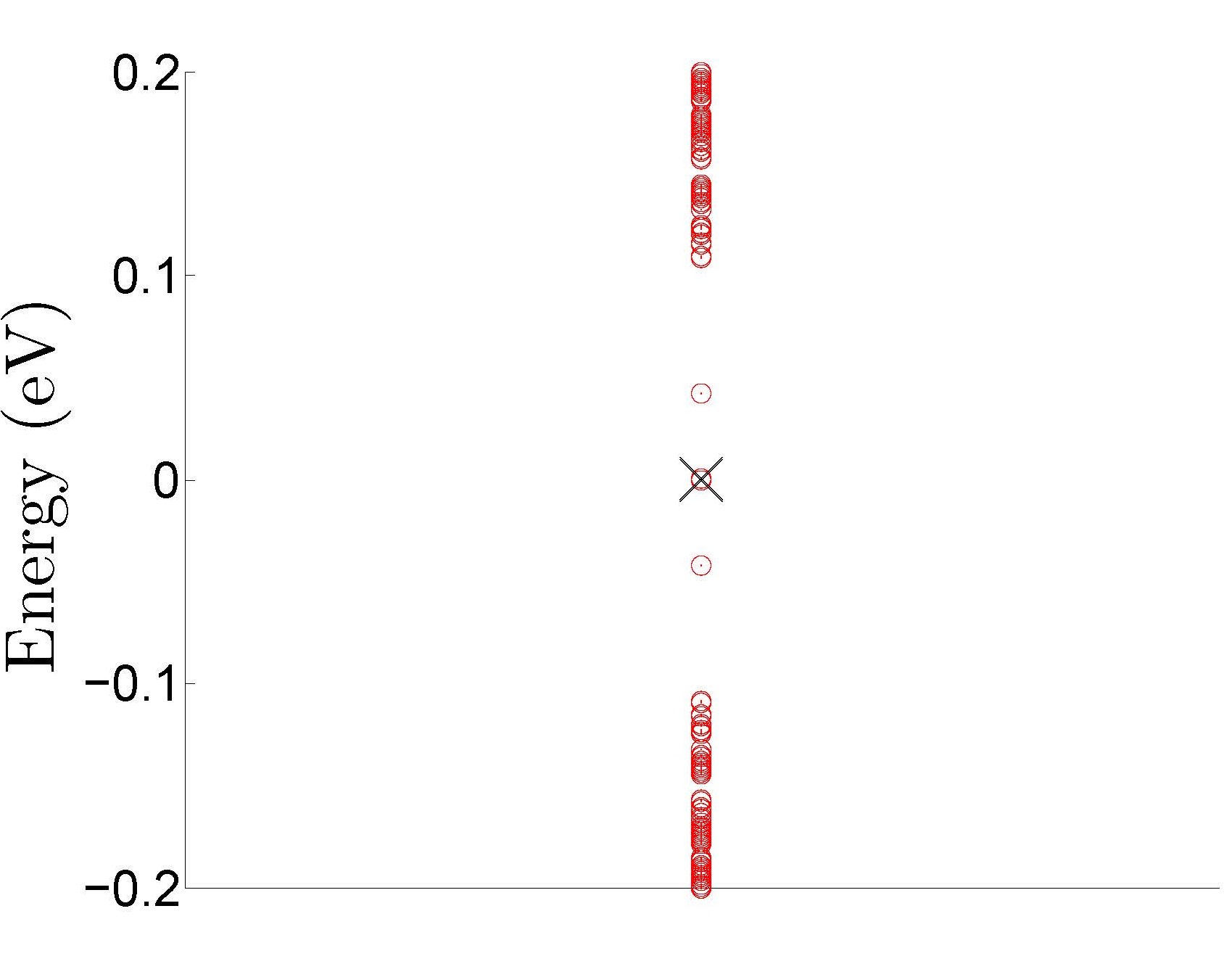}\;\includegraphics[width=0.24\textwidth]{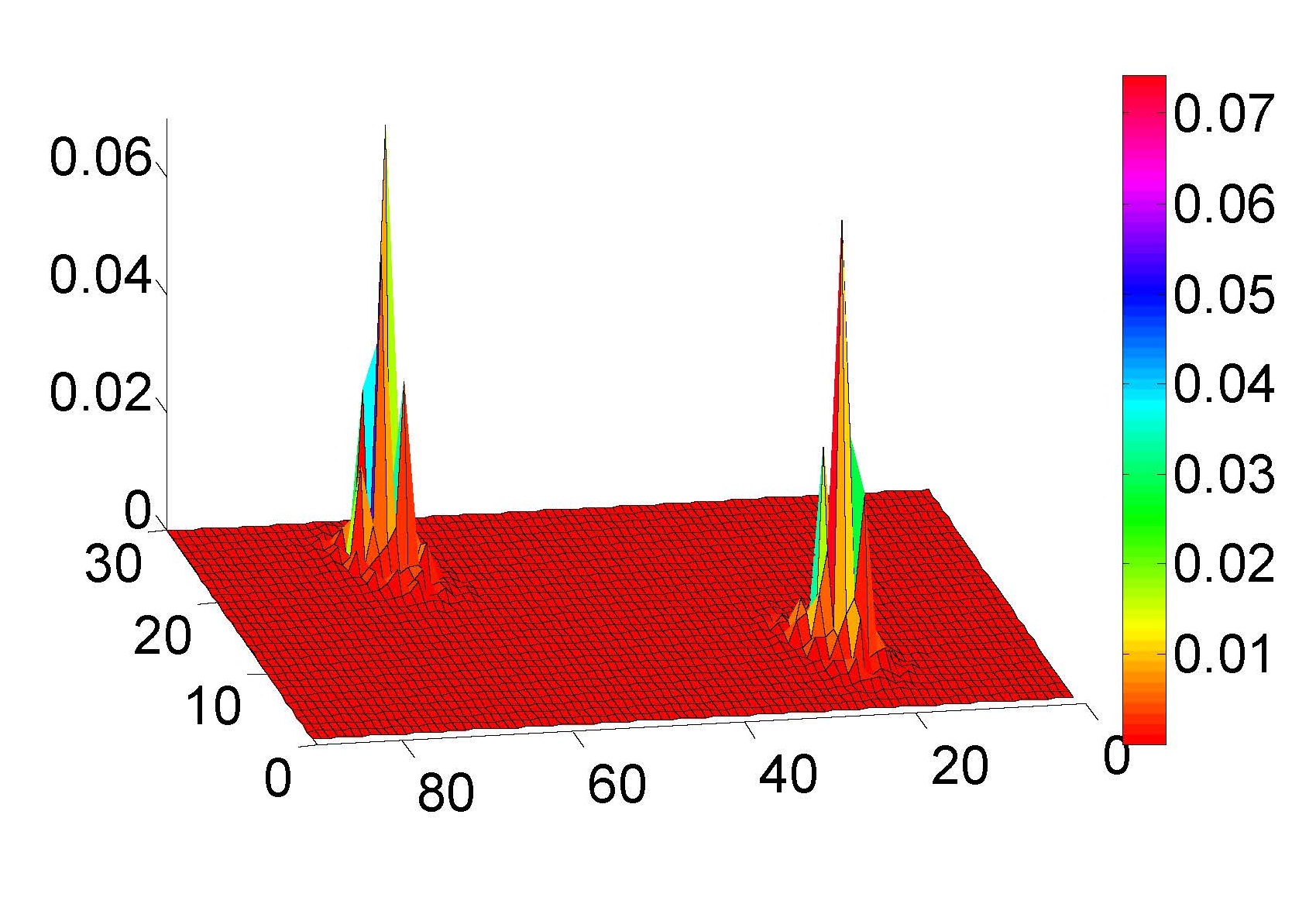}
\caption{(color online) Magnetic point defect energy spectrum and the MBS density profile on the triangular lattice. A pair of topologically-stable defects in the non-collinear magnetic order (NCMO) are located at
$(r_1,r_2)=(23.5,8.5)$ and $(68.5,23.5)$ on a $90\times30$ torus. The zero-energy ($E_0\approx0.0013~meV$) MBS (black cross) is separated from the first excited state by an energy $|E_1-E_0|\approx42~meV$. The electron density of the MBS is concentrated around the two defects. The bulk energy gap is about $100~meV$. The data are obtained for $\Delta_2=150~meV$ and $M=200~meV$ for the $1\times3$ noncollinear magnetic order.}\label{fig:defect}
\end{figure}

As discussed in the main text, the non-collinear magnetic order (NCMO) breaks the $SO(3)$ spin rotational symmetry and there exists topologically stable point defects of NCMO due to the nontrivial homotopy $\pi_1\big(SO(3)\big)=\mathbb{Z}_2$. Moreover a SC vortex in the Fu-Kane model \cite{Fu2008} is mapped to such a stable point defect of the NCMO in the topological superconductors discussed in this work. This suggests each stable point defect of the NCMO hosts a single MBS. Here we confirm this conclusion by an explicit calculation pf the point-defect energy spectrum.

We put two stable point defects in the NCMO configuration on a $L_1\times L_2$ torus generated by the following spatial pattern of the magnetic order parameter \footnote{The magnetic mass term on a site $(x,y)$ is given by $M_1(x,y)\sigma_x+M_2(x,y)\sigma_y$.} $M=M_1+\imth M_2$:
\begin{eqnarray}
M(x,y)=M_0e^{\imth2\pi x/3}e^{\imth\phi(x+\imth y)},~~~x,y\in\mathbb{Z},
\end{eqnarray}
where the phase angle $\phi(x+\imth y)$ is defined in (\ref{vortex pair angle}). $M_0$ is a constant complex number. Diagonalizing the BdG equation in the presence of the point-defect pair, we obtain the energy spectrum shown in Fig.~2(c) in the main text and reproduced here in Fig.~\ref{fig:defect}. Note that there is a zero-energy MBS within the gap, whose electron density is concentrated around the two point defects.

\section{C. Chirality of Majorana edge mode in the square lattice model}

\begin{figure}
 \includegraphics[width=0.5\textwidth]{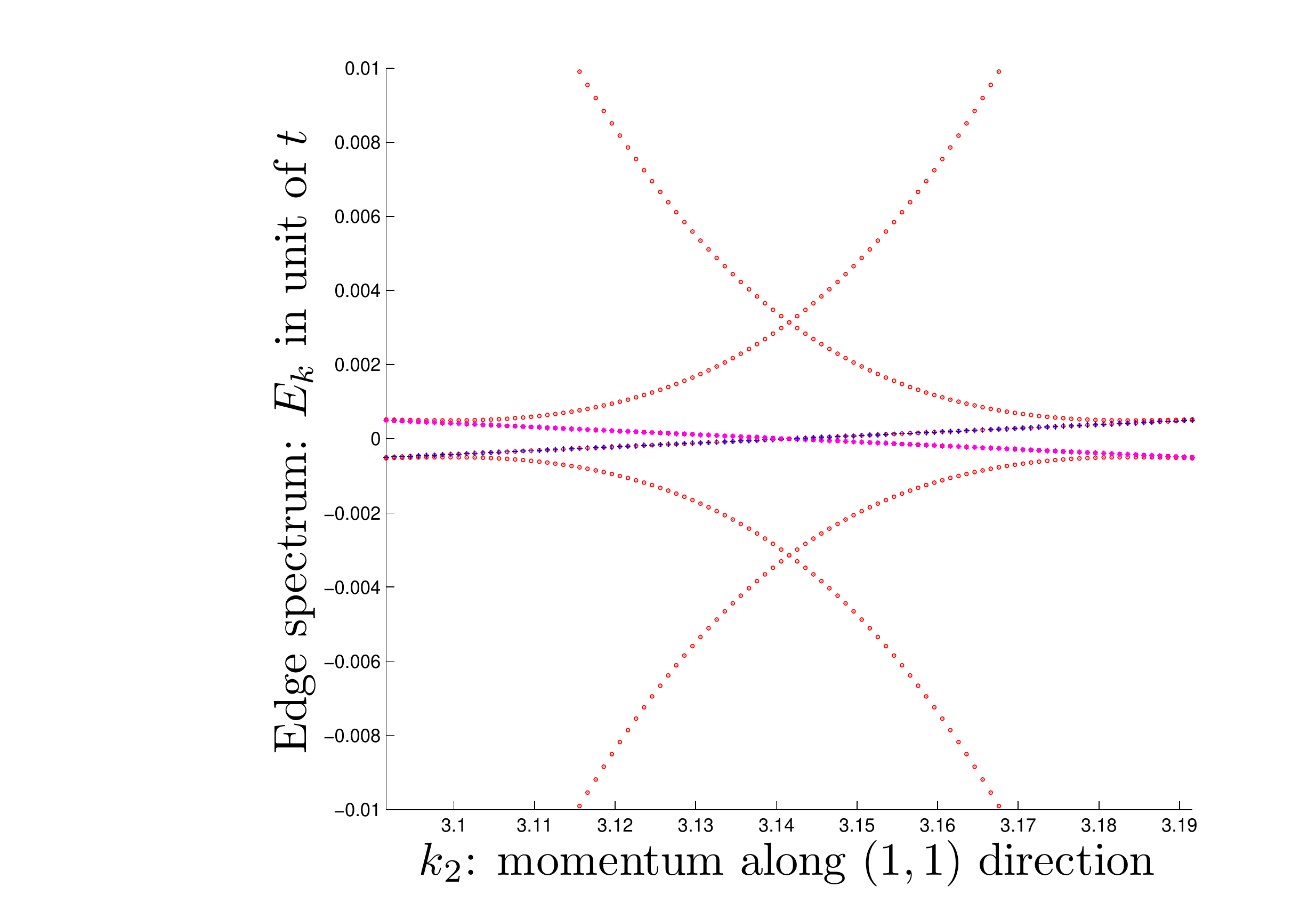}
\caption{(color online) Edge spectra as a function of momentum $k_2$ along the $(1,1)$ direction in the magnetic zone for a NN $d_{x^2-y^2}$ superconductor with $(\pi/2,\pi/2)$ non-collinear magnetic order. The spectrum is zoomed in around $k_2=\pi$ where narrow bands of edge mode are localized on the two edges (black and red), corresponding to the MBS. Magenta lines are the bulk states. A small positive NN imaginary pairing component $\Delta_{\tk,1}^{\prime\prime}=0.02 i \cos{k_1}$ is included.}\label{fig:1x4+}
\end{figure}

\begin{figure}
 \includegraphics[width=0.5\textwidth]{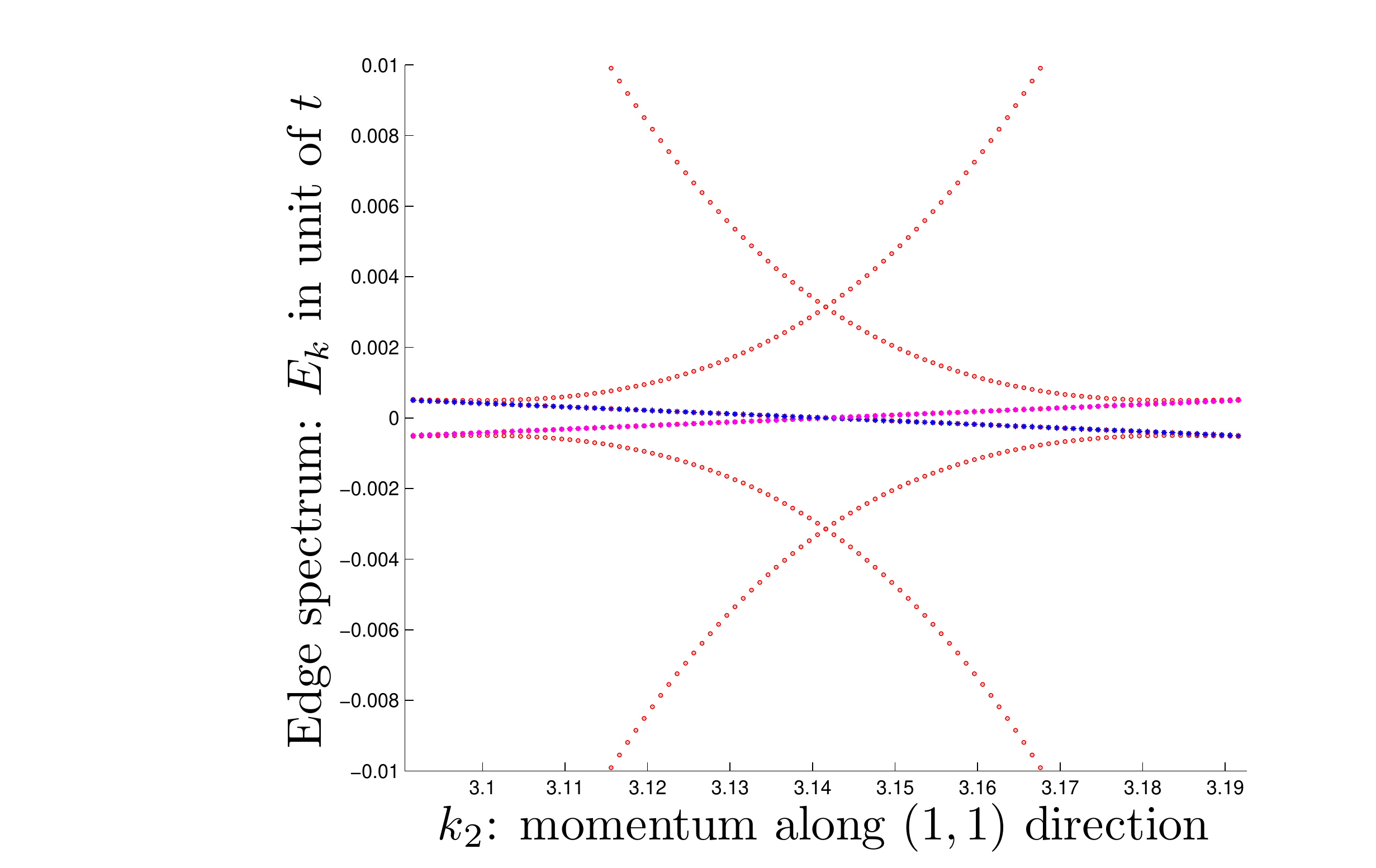}
\caption{(color online) Edge spectra as a function of momentum $k_2$ along the $(1,1)$ direction of the magnetic zone for a NN $d_{x^2-y^2}$ superconductor with $(\pi/2,\pi/2)$ non-collinear magnetic order. The spectrum is zoomed in around $k_2=\pi$ where narrow bands of edge mode are localized on the two edges (blue and red), corresponding to the MBS. Magenta lines are the bulk states. A small negative NN imaginary pairing component $\Delta_{\tk,1}^{\prime\prime}=-0.02 i \cos{k_1}$ is included.}\label{fig:1x4-}
\end{figure}

As discussed in the main text, when the pairing order parameter preserves time reversal symmetry 
there is a dispersionless MBS on the edge, as shown in Fig.~4 (main text) for the NN real $d_{x^2-y^2}$ pairing on the square lattice. However the presence of magnetic order breaks time reversal symmetry and a small imaginary pairing part in $\Delta_1$ can be induced. This would open up a gap in the bulk and result in a dispersing Majorana edge mode with a chirality $C=\pm1$. Here we demonstrate this phenomenon by calculating the edge spectrum of a complex $d_{x^2-y^2}$ pairing order parameter coexisting with $(\pi/2,\pi/2)$ non-collinear magnetic order with an additional small NN imaginary pairing component $\Delta_{\tk,1}^{\prime\prime}=\pm0.02\imth\cos{k_1}$. As shown in Figs.~\ref{fig:1x4+} and \ref{fig:1x4-}, the flat-band MBS in Fig.~4 (main text) begins to disperse in the presence of the imaginary pairing term. When the sign of this imaginary pairing is reversed, the chirality of the Majorana edge mode changes accordingly.

All the edge spectra in the square lattice case are obtained on a cylinder of size $L=4\times100=400$, with periodic boundary condition along the $(1,1)$ direction.


\end{document}